\documentclass{aastex631}

\usepackage{xcolor}

\usepackage{natbib} 
\usepackage{amsmath}
\usepackage{enumitem}
\shorttitle{Y6 Search}
\shortauthors{Bernardinelli et al.}

\defcitealias{Bernstein2000}{BK}
\defcitealias{Bernardinelli2019}{Paper I}

\newcommand{\E}{\mathrm{E}}

\newcommand{\des}{\textit{DES}}
\newcommand{\au}{\,\mathrm{au}}

\reportnum{DES-2021-0661}
\reportnum{FERMILAB-PUB-21-390-AE}

\begin{document} 

\title{A search of the full six years of the Dark Energy Survey for outer Solar System objects}

\author[0000-0003-0743-9422]{Pedro H. Bernardinelli}
\affiliation{Department of Physics and Astronomy, University of Pennsylvania, Philadelphia, PA 19104, USA}
\email{pedrobe@sas.upenn.edu}

\author[0000-0002-8613-8259]{Gary M. Bernstein}
\affiliation{Department of Physics and Astronomy, University of Pennsylvania, Philadelphia, PA 19104, USA}

\author[0000-0003-2764-7093]{Masao Sako}
\affiliation{Department of Physics and Astronomy, University of Pennsylvania, Philadelphia, PA 19104, USA}

\author{Brian Yanny}
\affiliation{Fermi National Accelerator Laboratory, P. O. Box 500, Batavia, IL 60510, USA}

\author{M.~Aguena}
\affiliation{Laborat\'orio Interinstitucional de e-Astronomia - LIneA, Rua Gal. Jos\'e Cristino 77, Rio de Janeiro, RJ - 20921-400, Brazil}
\author{S.~Allam}
\affiliation{Fermi National Accelerator Laboratory, P. O. Box 500, Batavia, IL 60510, USA}
\author{F.~Andrade-Oliveira}
\affiliation{Instituto de F\'{i}sica Te\'orica, Universidade Estadual Paulista, S\~ao Paulo, Brazil}
\affiliation{Laborat\'orio Interinstitucional de e-Astronomia - LIneA, Rua Gal. Jos\'e Cristino 77, Rio de Janeiro, RJ - 20921-400, Brazil}
\author{E.~Bertin}
\affiliation{CNRS, UMR 7095, Institut d'Astrophysique de Paris, F-75014, Paris, France}
\affiliation{Sorbonne Universit\'es, UPMC Univ Paris 06, UMR 7095, Institut d'Astrophysique de Paris, F-75014, Paris, France}
\author{D.~Brooks}
\affiliation{Department of Physics \& Astronomy, University College London, Gower Street, London, WC1E 6BT, UK}
\author{E.~Buckley-Geer}
\affiliation{Department of Astronomy and Astrophysics, University of Chicago, Chicago, IL 60637, USA}
\affiliation{Fermi National Accelerator Laboratory, P. O. Box 500, Batavia, IL 60510, USA}
\author{D.~L.~Burke}
\affiliation{Kavli Institute for Particle Astrophysics \& Cosmology, P. O. Box 2450, Stanford University, Stanford, CA 94305, USA}
\affiliation{SLAC National Accelerator Laboratory, Menlo Park, CA 94025, USA}
\author{A.~Carnero~Rosell}
\affiliation{Laborat\'orio Interinstitucional de e-Astronomia - LIneA, Rua Gal. Jos\'e Cristino 77, Rio de Janeiro, RJ - 20921-400, Brazil}
\author{M.~Carrasco~Kind}
\affiliation{Center for Astrophysical Surveys, National Center for Supercomputing Applications, 1205 West Clark St., Urbana, IL 61801, USA}
\affiliation{Department of Astronomy, University of Illinois at Urbana-Champaign, 1002 W. Green Street, Urbana, IL 61801, USA}
\author{J.~Carretero}
\affiliation{Institut de F\'{\i}sica d'Altes Energies (IFAE), The Barcelona Institute of Science and Technology, Campus UAB, 08193 Bellaterra (Barcelona) Spain}
\author{C.~Conselice}
\affiliation{Jodrell Bank Center for Astrophysics, School of Physics and Astronomy, University of Manchester, Oxford Road, Manchester, M13 9PL, UK}
\affiliation{University of Nottingham, School of Physics and Astronomy, Nottingham NG7 2RD, UK}
\author{M.~Costanzi}
\affiliation{Astronomy Unit, Department of Physics, University of Trieste, via Tiepolo 11, I-34131 Trieste, Italy}
\affiliation{INAF-Osservatorio Astronomico di Trieste, via G. B. Tiepolo 11, I-34143 Trieste, Italy}
\affiliation{Institute for Fundamental Physics of the Universe, Via Beirut 2, 34014 Trieste, Italy}
\author{L.~N.~da Costa}
\affiliation{Laborat\'orio Interinstitucional de e-Astronomia - LIneA, Rua Gal. Jos\'e Cristino 77, Rio de Janeiro, RJ - 20921-400, Brazil}
\affiliation{Observat\'orio Nacional, Rua Gal. Jos\'e Cristino 77, Rio de Janeiro, RJ - 20921-400, Brazil}
\author{J.~De~Vicente}
\affiliation{Centro de Investigaciones Energ\'eticas, Medioambientales y Tecnol\'ogicas (CIEMAT), Madrid, Spain}
\author{S.~Desai}
\affiliation{Department of Physics, IIT Hyderabad, Kandi, Telangana 502285, India}
\author{H.~T.~Diehl}
\affiliation{Fermi National Accelerator Laboratory, P. O. Box 500, Batavia, IL 60510, USA}
\author{J.~P.~Dietrich}
\affiliation{Faculty of Physics, Ludwig-Maximilians-Universit\"at, Scheinerstr. 1, 81679 Munich, Germany}
\author{P.~Doel}
\affiliation{Department of Physics \& Astronomy, University College London, Gower Street, London, WC1E 6BT, UK}
\author{K.~Eckert}
\affiliation{Department of Physics and Astronomy, University of Pennsylvania, Philadelphia, PA 19104, USA}
\author{S.~Everett}
\affiliation{Santa Cruz Institute for Particle Physics, Santa Cruz, CA 95064, USA}
\author{I.~Ferrero}
\affiliation{Institute of Theoretical Astrophysics, University of Oslo. P.O. Box 1029 Blindern, NO-0315 Oslo, Norway}
\author{B.~Flaugher}
\affiliation{Fermi National Accelerator Laboratory, P. O. Box 500, Batavia, IL 60510, USA}
\author{P.~Fosalba}
\affiliation{Institut d'Estudis Espacials de Catalunya (IEEC), 08034 Barcelona, Spain}
\affiliation{Institute of Space Sciences (ICE, CSIC),  Campus UAB, Carrer de Can Magrans, s/n,  08193 Barcelona, Spain}
\author{J.~Frieman}
\affiliation{Fermi National Accelerator Laboratory, P. O. Box 500, Batavia, IL 60510, USA}
\affiliation{Kavli Institute for Cosmological Physics, University of Chicago, Chicago, IL 60637, USA}
\author{J.~Garc\'ia-Bellido}
\affiliation{Instituto de Fisica Teorica UAM/CSIC, Universidad Autonoma de Madrid, 28049 Madrid, Spain}
\author{D.~W.~Gerdes}
\affiliation{Department of Astronomy, University of Michigan, Ann Arbor, MI 48109, USA}
\affiliation{Department of Physics, University of Michigan, Ann Arbor, MI 48109, USA}
\author{D.~Gruen}
\affiliation{Faculty of Physics, Ludwig-Maximilians-Universit\"at, Scheinerstr. 1, 81679 Munich, Germany}
\author{R.~A.~Gruendl}
\affiliation{Center for Astrophysical Surveys, National Center for Supercomputing Applications, 1205 West Clark St., Urbana, IL 61801, USA}
\affiliation{Department of Astronomy, University of Illinois at Urbana-Champaign, 1002 W. Green Street, Urbana, IL 61801, USA}
\author{J.~Gschwend}
\affiliation{Laborat\'orio Interinstitucional de e-Astronomia - LIneA, Rua Gal. Jos\'e Cristino 77, Rio de Janeiro, RJ - 20921-400, Brazil}
\affiliation{Observat\'orio Nacional, Rua Gal. Jos\'e Cristino 77, Rio de Janeiro, RJ - 20921-400, Brazil}
\author{S.~R.~Hinton}
\affiliation{School of Mathematics and Physics, University of Queensland,  Brisbane, QLD 4072, Australia}
\author{D.~L.~Hollowood}
\affiliation{Santa Cruz Institute for Particle Physics, Santa Cruz, CA 95064, USA}
\author{K.~Honscheid}
\affiliation{Center for Cosmology and Astro-Particle Physics, The Ohio State University, Columbus, OH 43210, USA}
\affiliation{Department of Physics, The Ohio State University, Columbus, OH 43210, USA}
\author{D.~J.~James}
\affiliation{Center for Astrophysics $\vert$ Harvard \& Smithsonian, 60 Garden Street, Cambridge, MA 02138, USA}
\author{S.~Kent}
\affiliation{Fermi National Accelerator Laboratory, P. O. Box 500, Batavia, IL 60510, USA}
\affiliation{Kavli Institute for Cosmological Physics, University of Chicago, Chicago, IL 60637, USA}
\author{K.~Kuehn}
\affiliation{Australian Astronomical Optics, Macquarie University, North Ryde, NSW 2113, Australia}
\affiliation{Lowell Observatory, 1400 Mars Hill Rd, Flagstaff, AZ 86001, USA}
\author{N.~Kuropatkin}
\affiliation{Fermi National Accelerator Laboratory, P. O. Box 500, Batavia, IL 60510, USA}
\author{O.~Lahav}
\affiliation{Department of Physics \& Astronomy, University College London, Gower Street, London, WC1E 6BT, UK}
\author{M.~A.~G.~Maia}
\affiliation{Laborat\'orio Interinstitucional de e-Astronomia - LIneA, Rua Gal. Jos\'e Cristino 77, Rio de Janeiro, RJ - 20921-400, Brazil}
\affiliation{Observat\'orio Nacional, Rua Gal. Jos\'e Cristino 77, Rio de Janeiro, RJ - 20921-400, Brazil}
\author{M.~March}
\affiliation{Department of Physics and Astronomy, University of Pennsylvania, Philadelphia, PA 19104, USA}
\author{F.~Menanteau}
\affiliation{Center for Astrophysical Surveys, National Center for Supercomputing Applications, 1205 West Clark St., Urbana, IL 61801, USA}
\affiliation{Department of Astronomy, University of Illinois at Urbana-Champaign, 1002 W. Green Street, Urbana, IL 61801, USA}
\author{R.~Miquel}
\affiliation{Instituci\'o Catalana de Recerca i Estudis Avan\c{c}ats, E-08010 Barcelona, Spain}
\affiliation{Institut de F\'{\i}sica d'Altes Energies (IFAE), The Barcelona Institute of Science and Technology, Campus UAB, 08193 Bellaterra (Barcelona) Spain}
\author{R.~Morgan}
\affiliation{Physics Department, 2320 Chamberlin Hall, University of Wisconsin-Madison, 1150 University Avenue Madison, WI  53706-1390}
\author{J.~Myles}
\affiliation{Department of Physics, Stanford University, 382 Via Pueblo Mall, Stanford, CA 94305, USA}
\affiliation{Kavli Institute for Particle Astrophysics \& Cosmology, P. O. Box 2450, Stanford University, Stanford, CA 94305, USA}
\affiliation{SLAC National Accelerator Laboratory, Menlo Park, CA 94025, USA}
\author{R.~L.~C.~Ogando}
\affiliation{Laborat\'orio Interinstitucional de e-Astronomia - LIneA, Rua Gal. Jos\'e Cristino 77, Rio de Janeiro, RJ - 20921-400, Brazil}
\affiliation{Observat\'orio Nacional, Rua Gal. Jos\'e Cristino 77, Rio de Janeiro, RJ - 20921-400, Brazil}
\author{A.~Palmese}
\affiliation{Fermi National Accelerator Laboratory, P. O. Box 500, Batavia, IL 60510, USA}
\affiliation{Kavli Institute for Cosmological Physics, University of Chicago, Chicago, IL 60637, USA}
\author{F.~Paz-Chinch\'{o}n}
\affiliation{Center for Astrophysical Surveys, National Center for Supercomputing Applications, 1205 West Clark St., Urbana, IL 61801, USA}
\affiliation{Institute of Astronomy, University of Cambridge, Madingley Road, Cambridge CB3 0HA, UK}
\author{A.~Pieres}
\affiliation{Laborat\'orio Interinstitucional de e-Astronomia - LIneA, Rua Gal. Jos\'e Cristino 77, Rio de Janeiro, RJ - 20921-400, Brazil}
\affiliation{Observat\'orio Nacional, Rua Gal. Jos\'e Cristino 77, Rio de Janeiro, RJ - 20921-400, Brazil}
\author{A.~A.~Plazas~Malag\'on}
\affiliation{Department of Astrophysical Sciences, Princeton University, Peyton Hall, Princeton, NJ 08544, USA}
\author{A.~K.~Romer}
\affiliation{Department of Physics and Astronomy, Pevensey Building, University of Sussex, Brighton, BN1 9QH, UK}
\author{A.~Roodman}
\affiliation{Kavli Institute for Particle Astrophysics \& Cosmology, P. O. Box 2450, Stanford University, Stanford, CA 94305, USA}
\affiliation{SLAC National Accelerator Laboratory, Menlo Park, CA 94025, USA}
\author{E.~Sanchez}
\affiliation{Centro de Investigaciones Energ\'eticas, Medioambientales y Tecnol\'ogicas (CIEMAT), Madrid, Spain}
\author{V.~Scarpine}
\affiliation{Fermi National Accelerator Laboratory, P. O. Box 500, Batavia, IL 60510, USA}
\author{M.~Schubnell}
\affiliation{Department of Physics, University of Michigan, Ann Arbor, MI 48109, USA}
\author{S.~Serrano}
\affiliation{Institut d'Estudis Espacials de Catalunya (IEEC), 08034 Barcelona, Spain}
\affiliation{Institute of Space Sciences (ICE, CSIC),  Campus UAB, Carrer de Can Magrans, s/n,  08193 Barcelona, Spain}
\author{I.~Sevilla-Noarbe}
\affiliation{Centro de Investigaciones Energ\'eticas, Medioambientales y Tecnol\'ogicas (CIEMAT), Madrid, Spain}
\author{M.~Smith}
\affiliation{School of Physics and Astronomy, University of Southampton,  Southampton, SO17 1BJ, UK}
\author{M.~Soares-Santos}
\affiliation{Department of Physics, University of Michigan, Ann Arbor, MI 48109, USA}
\author{E.~Suchyta}
\affiliation{Computer Science and Mathematics Division, Oak Ridge National Laboratory, Oak Ridge, TN 37831}
\author{M.~E.~C.~Swanson}
\affiliation{Center for Astrophysical Surveys, National Center for Supercomputing Applications, 1205 West Clark St., Urbana, IL 61801, USA}
\author{G.~Tarle}
\affiliation{Department of Physics, University of Michigan, Ann Arbor, MI 48109, USA}
\author{C.~To}
\affiliation{Department of Physics, Stanford University, 382 Via Pueblo Mall, Stanford, CA 94305, USA}
\affiliation{Kavli Institute for Particle Astrophysics \& Cosmology, P. O. Box 2450, Stanford University, Stanford, CA 94305, USA}
\affiliation{SLAC National Accelerator Laboratory, Menlo Park, CA 94025, USA}
\author{T.~N.~Varga}
\affiliation{Max Planck Institute for Extraterrestrial Physics, Giessenbachstrasse, 85748 Garching, Germany}
\affiliation{Universit\"ats-Sternwarte, Fakult\"at f\"ur Physik, Ludwig-Maximilians Universit\"at M\"unchen, Scheinerstr. 1, 81679 M\"unchen, Germany}
\author{A.~R.~Walker}
\affiliation{Cerro Tololo Inter-American Observatory, NSF’s NOIRLab, Casilla 603, La Serena, Chile}

\collaboration{1000}{(The DES Collaboration)}
\suppressAffiliations

\begin{abstract}
  We present the results of a search for outer Solar System objects in the full six years of data (``Y6'') from the Dark Energy Survey (\des). The \des\ covered a contiguous $5000\deg^2$ of the southern sky with $\approx 80,000$ $3\deg^2$ exposures in the $grizY$ optical/IR filters between 2013 and 2019. This search yielded 815 trans-Neptunian objects (TNOs), one Centaur and one Oort cloud comet, with 461 objects reported for the first time in this paper. We present the search methodology that builds upon our previous search carried out on the first four years of data (``Y4''). Here, all \des\ images were reprocessed with an improved detection pipeline that leads to an average completeness gain of 0.47 mag per exposure, as well as an improved transient catalog production and optimized algorithms for linkage of such detections into Solar System orbits. All objects reported herein were verified by visual inspection and by computing the
``sub-threshold significance'', namely the total signal-to-noise ratio in the stack of images in which the object's presence is indicated by the orbit fit, but no detection was reported.
This yields a highly pure catalog of trans-Neptunian objects complete to $r \approx 23.8$~mag and distances $29<d<2500$~au. The Y6 TNOs have  minimum (median) of 7 (12) distinct nights' detections and arcs of 1.1 (4.2) years, and will have $grizY$ magnitudes available in a further publication. We present publicly available software for simulating our observational biases that enable comparisons of population models to our TNO detections. Initial inferences demonstrating the statistical power of the DES Y6 catalog are:  the data are inconsistent with the CFEPS-L7 model for the classical Kuiper Belt;  the 16 ``extreme'' TNOs ($a>150$~au, $q>30$~au) are consistent with the null hypothesis of azimuthal isotropy; and non-resonant TNOs with $q>38$~au, $a>50$~au show a highly significant tendency to be sunward of the 
major mean motion resonances, as expected in models of resonance sweeping, whereas this tendency is not present for $q<38$~au. 
\end{abstract}

\section{Introduction} \label{sec:intro_y6}
The population of small bodies orbiting beyond Neptune is a remnant of events early in the formation of the Solar System. The current orbital distribution of these trans-Neptunian objects (TNOs) is the result of the migration of the giant planets \citep{Fernandez1984,Tsiganis2005,Levison2008} and, since the discovery of the {second} Kuiper belt object by \cite{Jewitt1993}, numerous subsequent surveys of the trans-Neptunian region have identified thousands of objects \citep[\emph{e.g.}][]{Jewitt1995,Gladman1998,Allen2001,Allen2002,Trujillo2001,Millis2002,Bernstein2004,Elliot2005,Fuentes2008,Fraser2010,Schwamb2010,Petit2011,Petit2017,Rabinowitz2012,Brown2015a,Bannister2016,Bannister2018,Sheppard2016,Sheppard2019,Weryk2016,Chen2018,Whidden2019,Bernardinelli2019}.\footnote{\cite{Bannister2020} presents a comprehensive review of these surveys and summarizes their discoveries.} The observed variety of dynamical classes \citep{Gladman2008} and surface compositions \citep{Brown2012} has led to different hypotheses about the formation of this region. Neptune's migration can trap planetesimals into mean motion resonances \citep{Malhotra1993,Malhotra1995}, and gravitational interactions between Neptune and these planetesimals can further excite their orbits \citep{Gomes2003}. More detailed models of the formation of this region include instabilities in Neptune's orbit \citep{Dawson2012,Batygin2012}, variations in Neptune's migration timescale and smoothness \citep{Nesvorny2015,Pike2017}, the effects of a potential unobserved giant planet in the outer Solar System \citep{Batygin2019}, close stellar encounters \citep{Pfalzner2018}, the birth environment of the Solar System \citep{Adams2010}, and Galactic tides \citep{Duncan2008}.

The Dark Energy Survey \citep[\des,][]{TheDarkEnergySurveyCollaboration2005,TheDarkEnergySurveyCollaboration2016} received an allocation of 575 observing nights on the 4m Blanco Telescope in Cerro Tololo between 2013 and 2019, with the primary objective of measuring the nature of the accelerated expansion of the universe and the spatial distribution of dark matter \citep{des_3x2,des_h0,des_bao,des_sn,PhysRevLett.122.171301}.   The $3\deg^2$, 520 Mpix Dark Energy Camera \citep[DECam,][]{Flaugher2015} was built for the survey, enabling
the ``wide'' component of the survey to image a contiguous $5000\deg^2$ area of the southern sky $\approx10$ times in each of the $grizY$ bands over 6 years. The interlaced ``supernova''  or ``deep'' survey, imaged a smaller area ($\approx30\deg^2$) in $griz$ at a $\approx$weekly cadence, aimed at detecting and characterizing Type Ia supernovae \citep{Bernstein2012}. \des\ has reported discoveries of individual TNOs of interest \citep{Gerdes2016,Gerdes2017,Becker2018,Khain2018,Lin2019}, as well as dynamical classifications for many detected objects \citep{TheDarkEnergySurveyCollaboration2016,Khain2020,Bernardinelli2019}, statistical analyses of the population of large-$a$, large-$q$ ``extreme'' TNOs \citep{Bernardinelli2020,Napier2021}, forecasts of future occultation events \citep{BandaHuarca2019} and a survey of machine learning techniques for TNO searches \citep{Henghes2020}. 

\citet[hereafter Paper I]{Bernardinelli2019} presented the results of a uniform TNO search in the first four years of \des\ {wide-survey} data (``Y4'') that yielded 316 objects. The Y4 search required the development of several algorithms for moving object identification, orbit linking and recovery, object confirmation, and simulations of object discovery and observational biases \citep[see also][]{Bernardinelli2020}. In this paper, we describe a search of the full six years of \des\ {wide-survey} data (``Y6'') for TNOs. Methodological improvements over the Y4 search are described in Section~\ref{sec:search}.  Section~\ref{sec:surveysim} characterizes the selection function for the Y6 \des\ TNO search, and describes tools that we make available for simulating \des\ discoveries given hypothetical TNO populations.
These tools are required for statistical comparisons of theoretical TNO populations to the Y6 catalog of \des\ TNOs.  We demonstrate that the detection efficiency for bound objects at 30--2000~au distance is almost entirely dependent on the mean $r$-band apparent magnitude of the TNO, and independent of light curve amplitude, color, or orbital elements, aside from the considerations of having to reside within the \des\ footprint for a sufficient fraction of the survey epoch.

Section~\ref{sec:catalog} presents the catalog of TNOs detected in the full \des\ wide survey data. This catalog has {817} confirmed objects (461 first discovered in this work) with $grizY$ photometry, high quality multi-year orbital solutions yielding precise dynamical classifications, and well characterized observational biases. This is the second largest TNO catalog from a single survey to date, as well as the largest catalog with multi-band photometry. Section \ref{sec:implications} discusses statistical comparisons between the \des\ objects and models of the trans-Neptunian region. Section \ref{sec:summary} summarizes this release. Appendix \ref{sec:triplets} describes a new triplet search algorithm developed for this search.


\section{The Year 6 search}
\label{sec:search}
The search for TNOs in the \des\ data has 3 major stages:
\begin{enumerate}
	\item The identification of single-night transients, that is, a source that appears in the sky in a given location for only a single night;
	\item Orbit linking, where we ``connect'' all sets of transients that could potentially come from the same outer-Solar-System object;
	\item Verification of the linkage through the ``sub-threshold significance'' (\texttt{STS}) statistic, whereby we stack all images containing the putative TNO that did \emph{not} yield a detection, thus yielding an independent check of whether the TNO actually exists.
\end{enumerate}

These steps are summarized below, with more detailed description of any aspects that have changed since the Y4 search. Table \ref{tb:stages_y6} outlines the search, and presents the number of sources relevant to each step of the processing.

\begin{deluxetable}{lccc}
	\tabletypesize{\footnotesize}
  \tablecaption{Stages of the Y6 processing. The entries in \emph{italics} indicate each intermediate step, going from the SE catalogs to the final catalog of TNOs. The columns also indicate the number of elements and the number of injected elements (when appropriate) at each stage of the processing. The final column describes changes, if present, between this step and the Y4 procesing. The search required between 15 and 20 million CPU hours to be completed.\label{tb:stages_y6}}
	\tablehead{\colhead{Catalog/\textit{Processing step}} & \colhead{No. of real elements} &
        \colhead{No. of injected elements} & \colhead{Changes in Y6}}
	\startdata
	Single epoch detections (\textsection \ref{sec:desdata})& $1.60 \times 10^{10} $ & $1.10\times10^5$ &
        Lower threshold \\
	$\rightarrow$\emph{Blinded fake injection} (\textsection \ref{sec:fakes}) $\rightarrow$ &  &  & Changed order \\
	$\rightarrow$\emph{Transient identification} (\textsection \ref{sec:transients}) $\rightarrow$ &  &  & Changed order  \\
	$\rightarrow$\emph{Coadd avoidance} (\textsection \ref{sec:coaddavoid}) $\rightarrow$ &  &   & New\\
	$\rightarrow$\emph{Pixel-level masking} (\textsection \ref{sec:pixmask}) $\rightarrow$ &  &  & New \\
	Transients (\textsection \ref{sec:transientcat}) & $1.08\times10^8$  & $1.05\times10^5$   \\
	$\rightarrow$\emph{Pair finding} (\textsection \ref{sec:linking}) $\rightarrow$ &  & \\
	Pairs & $ \gtrsim 10^{13}$ & \nodata \\
	$\rightarrow$\emph{Triplet finding (\textsection \ref{sec:triplets})}$\rightarrow$ & & & Modified \\
	Triplets & $\approx 10^{12}$  & \nodata &  \\
	$\rightarrow$\emph{Orbit growing (\textsection \ref{sec:linking})}$\rightarrow$ & & \\
	$\rightarrow$\emph{Fake unblinding} $\rightarrow$ & & \\
	Sevenlets & 31064 &  \\
	$\rightarrow$\emph{Reliability cuts (\textsection \ref{sec:linking})}$\rightarrow$ & & \\ 
	Candidates & 9081 & 3937 \\ 
	$\rightarrow$\emph{Sub-threshold significance test (\textsection \ref{sec:sts})}$\rightarrow$ & & & Lower threshold \\
	Visual inspection & 872 & \nodata \\
	Confirmed objects & 817 & \nodata
	\enddata
\end{deluxetable}

\subsection{Data acquisition and image processing}
\label{sec:desdata}
The search presented here uses the \des\ Y6A2 internal release. {Of 83,706 exposures processed from the wide survey, 76,217 pass quality cuts and their detections are entered into the ``single-epoch'' (SE) catalog. ``Coadd''  images are created as the average (per band) of all exposures in a given region of sky, and the detections and measurements derived from these images comprise the coadd catalog.} The \des\ image processing pipeline is described in detail in \cite{Morganson2018}, and the coadd catalogs as well as images used in this work correspond to those in the \des\ Data Release 2 \citep{DESDR2}. 

The nominal survey strategy \citep{Diehl2016,Diehl2018} was such that the $5000\deg^2$ footprint was tiled with $10\times 90 \, \mathrm{s}$ exposures in each of the $griz$ bands (and $6\times 45 + 2\times 90 \, \mathrm{s}$ exposures in the $Y$ band).  Each point within the footprint has been imaged by working detector pixels in 7--10 exposures per band, typically 8 \citep{DESDR2}. The observation scheduler \citep{obstac,Neilsen2019} specifically \emph{avoids} repeated exposures in the same region and in the same night, except for successive exposures with the same pointing in different bands, so the motion of a TNO is not readily detectable in our data. We have elected to remove the $Y$ band catalogs from the search, since most TNOs have very low $S/N$ in this filter. We do, however, use these images for $Y$-band photometric measurements of any TNO discovered in the $griz$ data, and for additional astrometric measurements of bright objects that have $Y$ band detections. The $gizY$ magnitudes of the objects reported herein will be presented in a future publication.

Each \des\ exposure has its astrometric solution mapped to \emph{Gaia} DR2 \citep{GaiaDR2}. The DECam astrometric model is fully described in \cite{Bernstein2017astro}, and all fixed optical distortions are known to $\approx 1\, \mathrm{mas}$ RMS.
{This astrometric model includes two color-dependent effects---differential chromatic refraction in the atmosphere, and lateral color in the DECam prime-focus corrector.  For the TNO search we use astrometric positions that assume a nominal mean TNO color for all detections.  Once the TNOs are linked and validated, we can determine a $g-i$ color for each.  We then recalculate the positions using the measured color, and re-fit the orbit.}
The dominant astrometric error for the brightest TNO (Eris) is from turbulent atmospheric distortions, with a typical RMS value of $7 \, \mathrm{mas}$ \citep{Fortino2020}.  All other TNOs have astrometric accuracy limited by shot noise in the centroiding.  {The estimations of these measurement uncertainties are shown to be accurate to $\lesssim 10\%$ by the $\chi^2$ values per degree of freedom of the orbit fits for Eris (62.9/58) and for the entire final catalog (28832/25236), that is, there is no overfitting in the data.}


The \des\ detections are calibrated to a photometric system that is highly uniform across the focal plane \citep{Bernstein2017phot}
and across all exposures \citep{Burke2017}, as demonstrated by an RMS difference from the \emph{Gaia} DR2 catalog of only $2.5\, \mathrm{mmag}$ \citep{DESDR2}. The photometric measurements reported for our TNOs are obtained using a dedicated photometric pipeline, whose details will be presented in a future publication. We report the average PSF-fitting flux over all images of a TNO in a given band, with the flux of each individual detection scaled to a nominal distance for that object. 

\subsection{Detection threshold optimization \& characterization}
\label{sec:det_thresh}
We have improved the image processing pipeline presented in \cite{Morganson2018} to allow fainter sources to be detected in each exposure. This is done by lowering the \textsc{Source Extractor} \citep{Bertin1996} detection threshold to $\mathtt{DETECT\_THRESH} = 0.8$ instead of the previous 1.5 used in the \des\ First Cut \citep{Morganson2018} processing, and using a detection kernel that better matches the typical point spread function (PSF), changing from a boxcar \citep{Morganson2018} to a Gaussian kernel. To optimize the threshold, we compared the catalogs of sources detected in \des\ exposure to the significantly deeper Hyper Suprime-Cam Subaru Strategic Program \citep[HSC;][]{Aihara2018} XMM-LSS catalog from their first data release. By using HSC, we have not only an independent set of measurements (so single-night transients from any \des\ exposure are \emph{not} present on the HSC images), but also guarantee completeness of the reference catalog to fainter magnitudes than any \des\ exposure. The \des\ sources that do not match HSC detections correspond to astrophysical transients and/or to spurious detections coming from noise fluctuations or defects in the image. Any lower choice of \texttt{DETECT\_THRESH} leads to a substantial increase in transient density, clearly due to detection of noise fluctuations, making it infeasible to conduct the search despite possible small gains in completeness for TNOs.

Once all \des\ images have been cataloged using the new Y6 detection parameters, we evaluate the detection threshold for unresolved sources as follows. For a given \des\ single exposure, we find all of the unresolved sources in the coadd catalog that overlap the exposure, and see whether each appears in the SE catalog.
The list of (un)matched sources is used to constrain a 
completeness function for each \des\ exposure. The completeness model is defined by the magnitude of 50\% completeness $m_{50}$, a scaling factor $c$ and transition sharpness $k$ such that the probability for a source with magnitude $m$ be detected in this exposure is given by
\begin{equation}
	p(m) = \frac{c}{1 + \exp(k(m - m_{50}))} . \label{eq:probdet}
\end{equation}

\begin{figure}[ht!]
	\centering
	\includegraphics[width=0.49\textwidth]{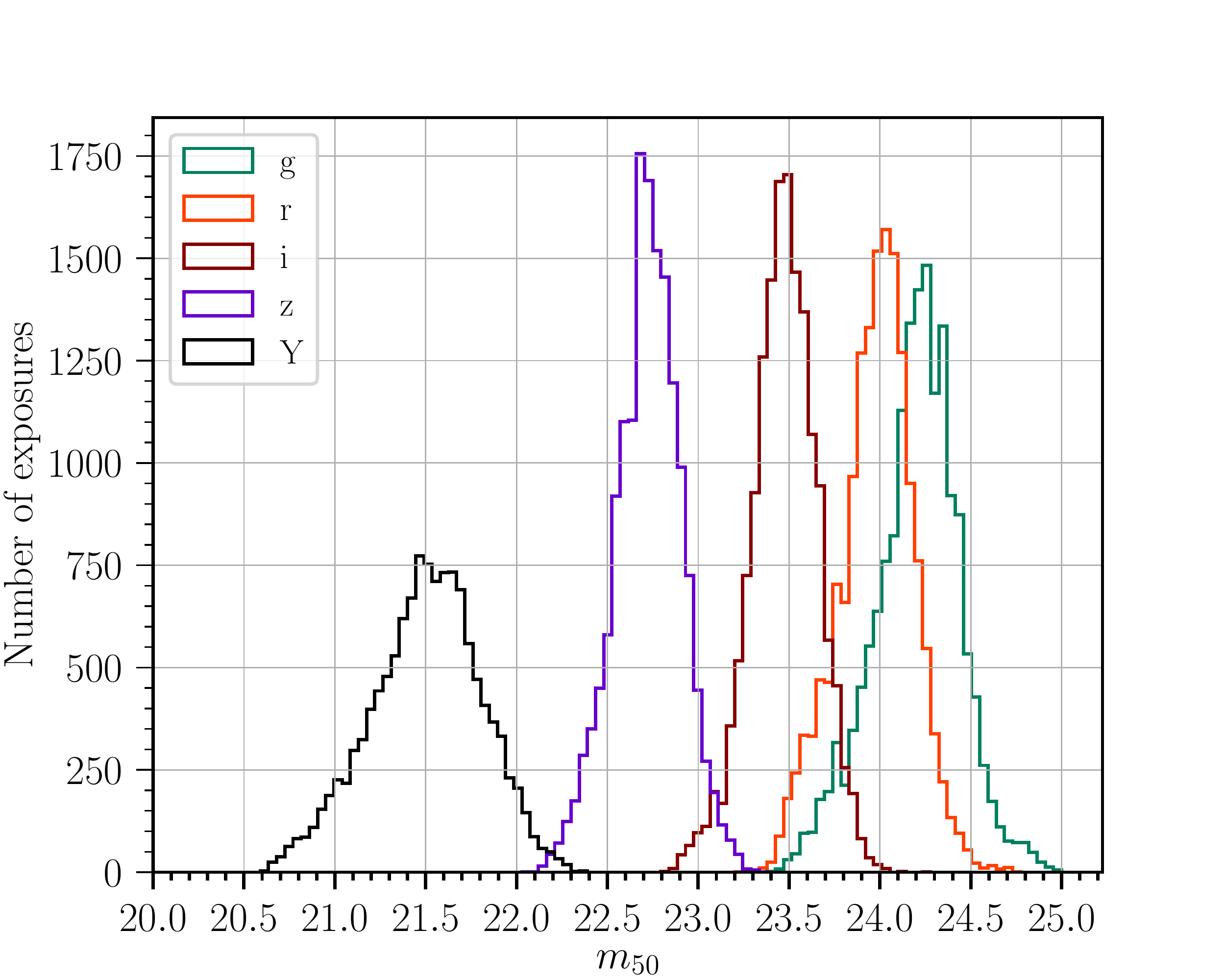}
	\includegraphics[width=0.49\textwidth]{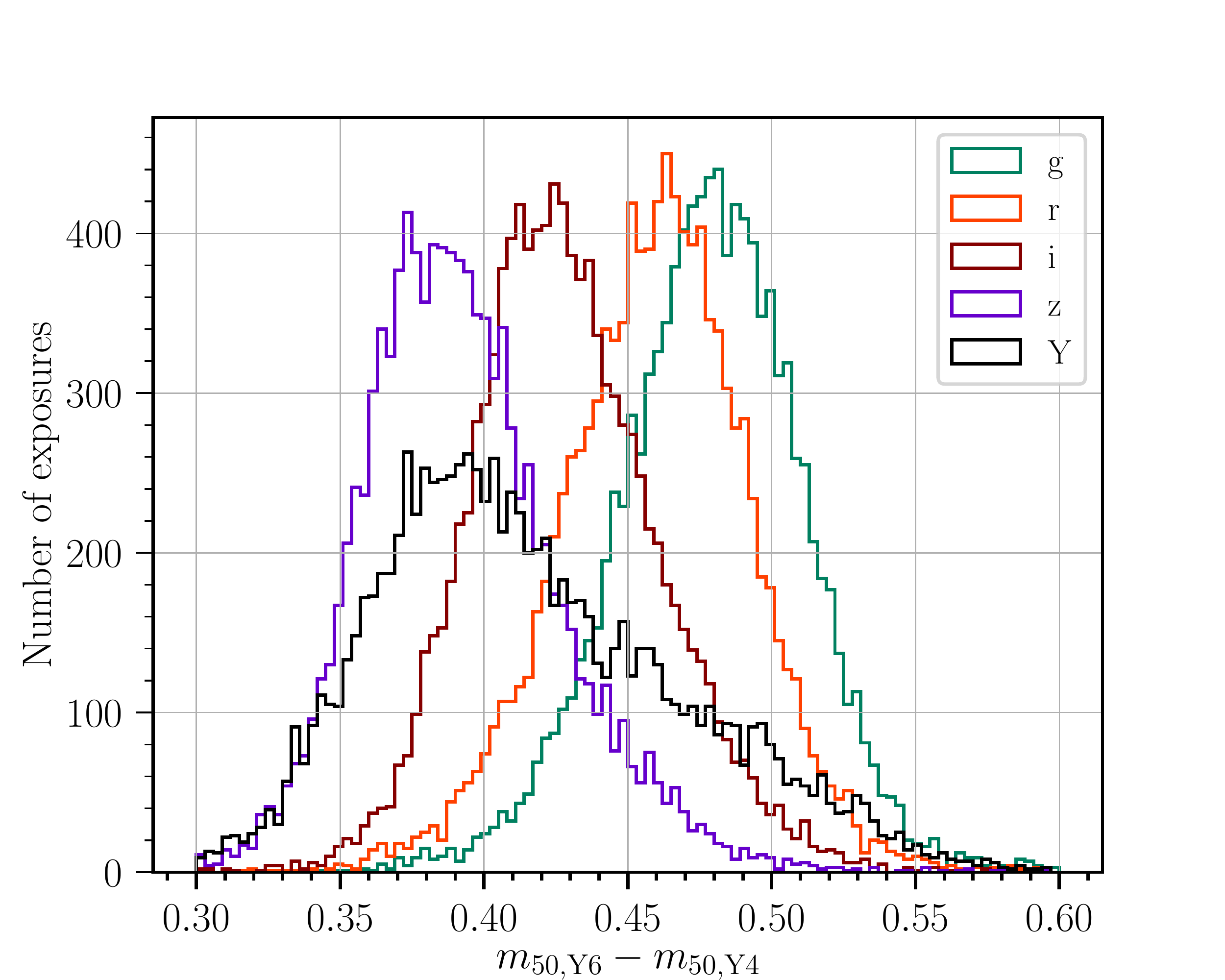}
	\caption{\emph{Left:} Distribution of single-exposure detection limit $m_{50},$ as defined in Eq.~(\ref{eq:probdet}), for all the Y6 exposures in each band. \emph{Right:} Histogram of difference in $m_{50}$ for between the year 4 and year 6 processing for each exposure. The mean gain in depth is $\sim0.46$ for $r$ band. This gain comes from primarily from the optimized detection threshold, and to a lesser extent from the deeper \des\ Y6 coadd catalogs used to veto static sources. \label{im:m50change}}
\end{figure}

{The left-hand side of Figure~\ref{im:m50change} plots histograms of the $m_{50}$ values derived for all of the accepted Y6 exposures.  The median depths are {24.2,  24.0,  23.5, 22.7, and 21.5~mag} in the $grizY$ bands, respectively.  These are a significant improvement over the depths attained in the Y4 processing, even for the same images.}  
Figure \ref{im:m50change} shows the histogram of changes in $m_{50}$ between the Y4 and Y6 processing for all exposures contained in both. An average of $\sim 0.47 \, \mathrm{mag}$ is gained in depth in $r$ band exposures.

\subsection{Injection of synthetic TNOs}
\label{sec:fakes}
We inject a synthetic population of TNOs (``fakes'') into the SE catalog in order to test for transient recovery and linking efficiencies.  Unlike in \citetalias{Bernardinelli2019}, this synthetic population is injected \emph{before} the production of the transient catalog, {allowing the simulation to measure the (in)efficiency of the transient-selection process as well as those of the linking process and selection cuts.}

Similar to \citetalias{Bernardinelli2019}, we construct an ensemble of orbits isotropically distributed in the Cartesian coordinates $\{x_0,y_0,z_0,\dot{x}_0,\dot{y}_0,\dot{z}_0\}$ for heliocentric distances $25 \, \mathrm{au} < d < 2500 \, \mathrm{au}$ and velocities $0 \leq v \leq v_\mathrm{esc}(d)$ so as to cover a wide range of sky locations, semi-major axes and eccentricities. We construct a sample of $\approx5000$ fakes intersecting the \des\ observations, from three different populations:
\begin{itemize}
	\item 80\% of the fakes are uniformly distributed in barycentric distance between $25$ and $60\,\mathrm{au}$, with the angular positions and velocities isotropically distributed using the method of \citetalias{Bernardinelli2019};
	\item 10\% of the population is logarithmically distributed between $60$ and $2500 \, \mathrm{au}$;
	\item The remaining 10\% are generated by constructing a larger isotropic population between $25$ and $60 \, \mathrm{au}$, but keeping only those with inclination $i < 20\degr$.
\end{itemize}
We have elected to include a larger number of objects at lower distances or inclinations than in \citetalias{Bernardinelli2019}, as these are more representative of the real trans-Neptunian population, {and larger numbers of synthetic TNOs are needed in these regimes to keep the shot noise of the simulations subdominant to the shot noise of the real detections.}

\begin{figure}[ht!]
	\centering
	\includegraphics[width=0.7\textwidth]{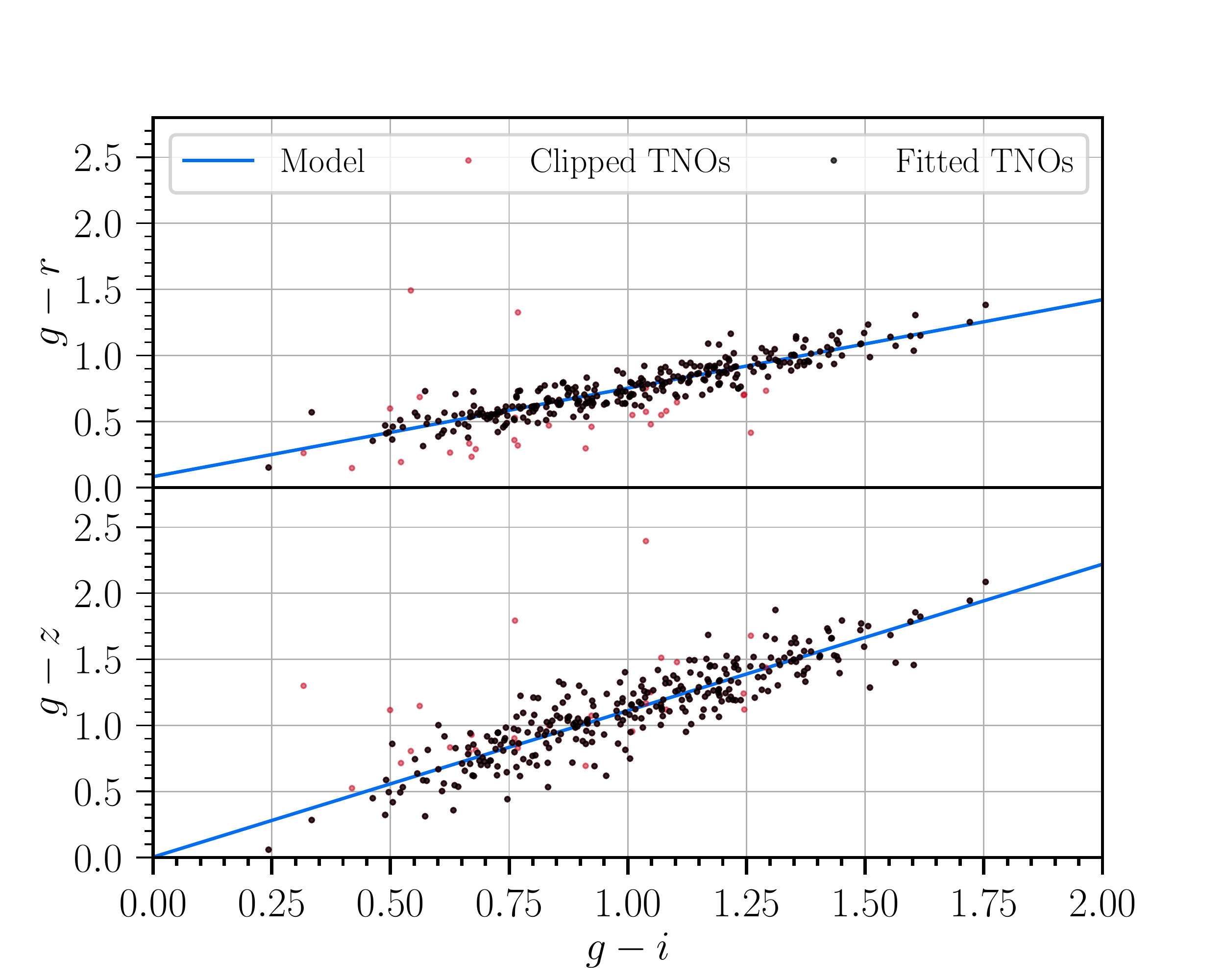}
	\caption{Trend lines for the $g - r$, $g - i$ and $g - z$ colors for the Y4 sample (black points), as well a linear regression (blue lines) for this data. Objects which were clipped are shown in red. We find that $\{\alpha_b,\beta_b\}$ is $\{1.49, -0.12\}$ for $i$ band and $\{1.65, -0.133\}$ for the $z$ band.\label{im:colors}}
\end{figure}

We also generate for each fake a mean $r$-band magnitude uniformly distributed in $20 \leq m_r \leq 24.5,$ and a color $0.4 < g - r < 1.5$. From this color index, we generated the other colors ($g - i$ and $g - z$) by a linear parametrization for band $b$
\begin{equation}
	g - b = \alpha_b (g - r) + \beta_b,\label{eq:colors}
\end{equation}
with the values for $\alpha_{b}$ and $\beta_{b}$ found by fitting to the observed colors of the 316 TNOs from the Y4 search (see Figure \ref{im:colors}). The objective of this fit is \emph{not} to measure correlations of TNO colors, rather just to get representative color variation into the fakes. So we do not include measurement errors while fitting, and we apply a sigma-clipping algorithm to remove outliers that are more than $3\sigma$ away from the fits in any of the three dimensions. The fitted color trends are shown in Figure \ref{im:colors}.

Differences in the surface reflectance as well as in the shape of TNOs can introduce apparent-magnitude variability on time scales shorter than the \des\ cadence \citep[see][and references therein]{Alexandersen2019}. To simulate this, each object also is assigned one of three possible light curve scenarios:
\begin{itemize}
	\item No light curve (constant magnitude, 50\% of the sample);
	\item Light curve with variation $\delta m \equiv A \sin\varphi$ and amplitude $A = 0.2$ (25\%);
	\item Light curve with the same variation as before and $A = 0.5$ (25\%).
\end{itemize}
These amplitudes were chosen to represent a wide range of TNO variability \citep{Alexandersen2019},
with peak-to-peak excursions of up to 1~mag across the $30$ to $50$ potential detections of each fake. Since the {time between \des\ observations of any given object is usually many times longer than a typical TNO rotation period,}
we do not {assign a period to each object,} but rather, for each observation, we draw a random phase $\varphi \in [0,2\pi)$. 

{For each observation of each fake TNO, we draw a uniform deviate $0<r<1$, and use Eq.~(\ref{eq:probdet}) to find a probability of detection.  If $p(m+\delta m) \ge r$ (where $\delta m$ is simulated with a random phase for each exposure),}
the synthetic source is considered as detected and is added to the SE catalog. Finally, each detected source receives an astrometric shot-noise error $\sigma_a$. For a given flux $f$, a background limited source has $\sigma_a \propto f^{-1}$, while for brighter sources $\sigma_a \propto f^{-1/2}$. Since most sources of interest are faint, we fixed $\sigma_a \propto f^{-1}$, and used the Y4 transient detections for magnitudes between $20 < m < 22$ for each band to find the expected amplitude of this shot noise. This corresponds to $\sigma_a(m_b = 20) = \{5.8, 4.9, 6.5, 9.6\} \, \mathrm{mas}$ for $b = \{g,r,i,z\}$, respectively. These shot-noise errors then get added in quadrature to each exposure's atmospheric turbulence covariance matrix \citepalias[see section 2.3 of][]{Bernardinelli2019}, and each object has its nominal positions shifted by drawing from the two-dimensional Gaussian defined by this sum of covariance matrices.

Section \ref{sec:surveysim} presents a discussion of the survey completeness versus the color, light-curve, and orbital characteristics of the fake TNOs.

\subsection{Transient identification}
\label{sec:transients}
The transient identification algorithm is the same as the one presented in \citetalias{Bernardinelli2019}. We begin by matching all detected SE and coadd sources by using a friends-of-friends (FoF) algorithm that links detections within $0.5\arcsec$ of each other, and, for each output group, we ask whether or not this source is matched to a coadd detection ($\mathtt{COADD} = 1,0$).  Next we evaluate how far apart in time are the first and the last detections of this match group ($\Delta t \equiv t_\mathrm{last} - t_\mathrm{first}$). Then, for the groups that have both a coadd and a SE source, we ask what is the least negative deviation in magnitude between a SE detection and the coadd in the same band, $\Delta m \equiv m_\mathrm{SE} - m_{\mathrm{coadd}}$. 

The coadd image at the location of a SE detection of a solar system object has this object's flux
in only one exposure out of $K$ in the same band and location, that is, the coadding process reduces the flux of a solar system object by a factor $1/K$, corresponding to a $\Delta m \lesssim - 1.2$ for $K \geq 3$. We include in the transient catalog detections that satisfy
\begin{itemize}
	\item $\mathtt{COADD} = 1$, $\Delta t < 2$ days, $\Delta m \leq -1.2$; \textsc{OR} 
	\item $\mathtt{COADD} = 0$, $\Delta t < 2$ days.
\end{itemize}
This process yields 160 million transient candidates, with the injections of fake TNOs detected with efficiency of 98.6\%.

After these transients are identified, the first cleaning step is to remove large clusters of spurious detections.  We apply a FoF algorithm with a linking radius of $30\arcsec$ to the catalog of all transients, and discard groups with over 20 detections, or groups with between 10 and 20 members whose detections have a tendency to lie along a line or a curve \citepalias[see section 2.4 of][]{Bernardinelli2019}, masking 17 million transients (11\%) and only 0.2\% of the fakes.

\subsection{Coadd avoidance radius}
\label{sec:coaddavoid}
Static sources {with complex shapes and/or low $S/N$ can be split into multiple detections in ways that differ from exposure to exposure.  This can generate}
SE detections that spread in radius more than the $0.5\arcsec$ used for the friends-of-friends matching.  Such detections can then be mistakenly identified as transients.

\begin{figure}[h!]
	\centering
	\includegraphics[width=0.49\textwidth]{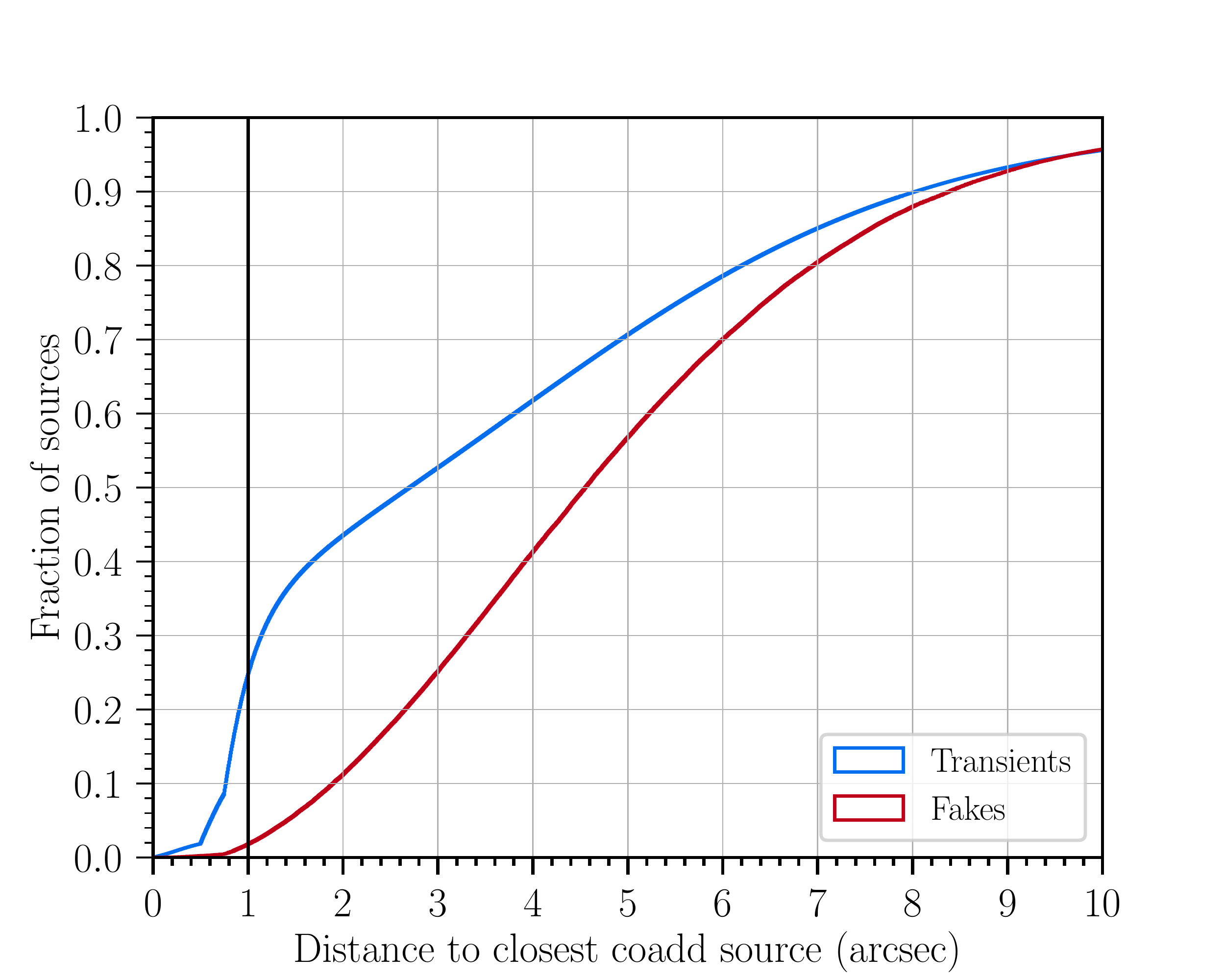}
	\caption{Cumulative histogram of distance between each identified transient and the closest coadd source, measured for all putative transients (blue), and for the fake detections injected in the catalog (red). The blue curve shows a steep growth for distances $\lesssim1\arcsec$, while the fakes, which are uniformly distributed over the footprint, show much lower fractions with close coadds---$24.3\%$ ($1.8\%$) of the transients (fakes) are closer than $1\arcsec$ to a coadd source.  {This indicates that the close pairs are primarily associated with static sources, and not true solar-system transients, so we remove them from the transient catalog.} 
       \label{im:distcoadd}}
\end{figure}

Figure \ref{im:distcoadd} shows in blue the cumulative histogram of the distance between each transients and its closest coadd source {(for those with $\texttt{COADD}=0$).}  The red line shows the same quantity for the injected fakes, i.e. a truly randomly placed population.  The significant excess of the former at low radii indicates that, for distances less than $\approx1\arcsec$, the putative transients are not independent of the static coadd sources, and contamination such as hypothesized above is present.

We therefore exclude from the transient catalog any SE detection that is within
a ``coadd avoidance radius'' of $1\arcsec$ of a coadd catalog source source, but had $\mathtt{COADD}=0$.
This process removes 32 million ($24.3\%$) of all identified transients, and only $1.8\%$ of the fakes. This percentage has a weak dependence on galactic latitude $b$, increasing to $2.1\%$ within $|b| \sim 20\degr$, and dropping to $1.5\%$ for sources farther than $|b| > 70\degr$.

\subsection{Pixel-level masks}
\label{sec:pixmask}
\begin{figure}[h!]
	\centering
	\includegraphics[width=0.8\textwidth]{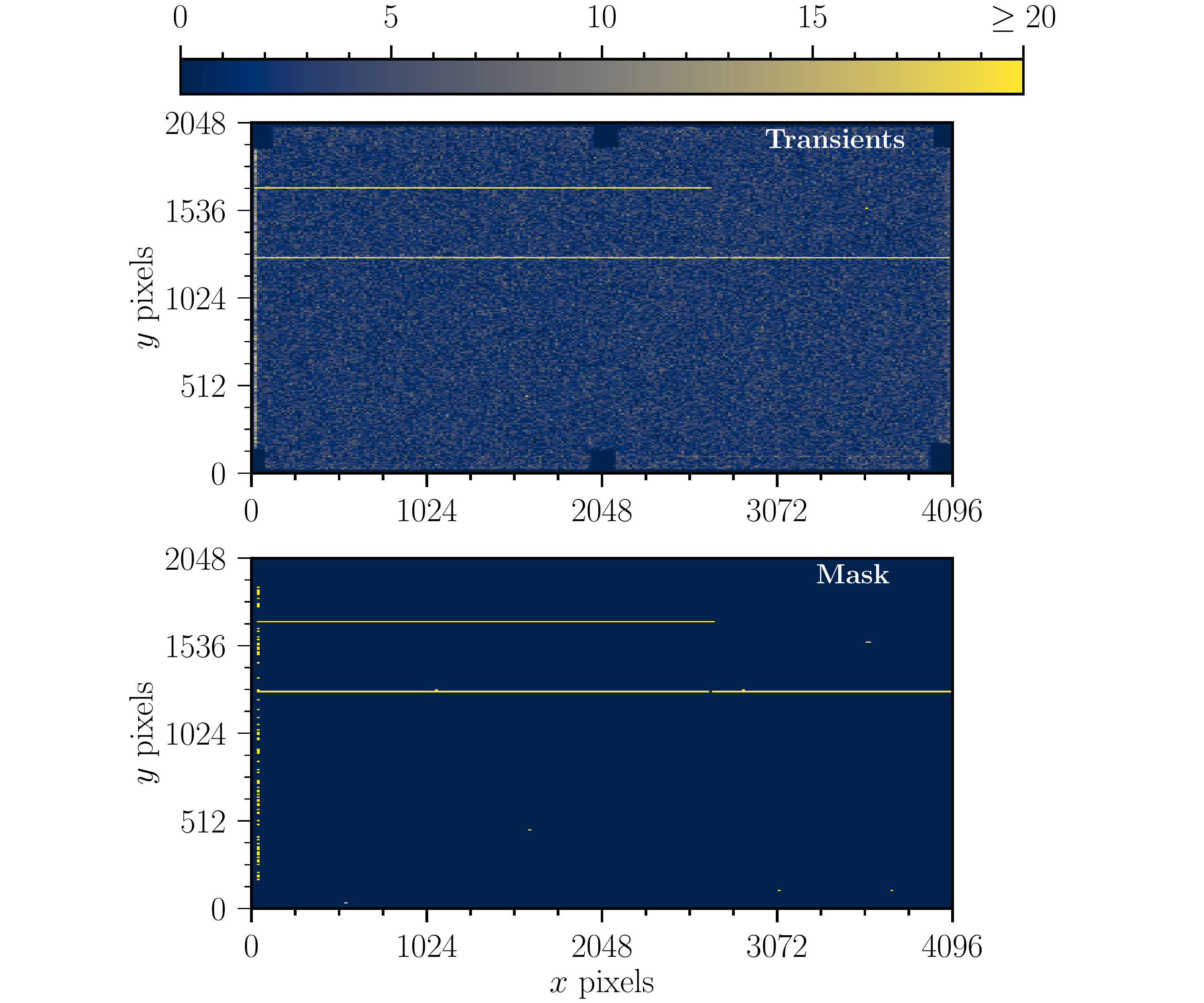}
	\caption{Example of the transient density per $8 \times 16$ pixel bin in a CCD, with the threshold for masking set to 10 transients per bin. The bottom panel shows the location of the masks in yellow, while the blue region is the unmasked portion of the CCD. A few features can be seen here, with an excess of transients (and, therefore, of masked regions) in the low-$x$ pixel counts of the CCD, as well as two bad columns in the CCD.\label{im:ccdmask}}
\end{figure}

A final masking process has been applied to the transient catalog by combining the data from each CCD in an observing ``epoch'' \citep[roughly two epochs per year, see Table 2 of][]{Morganson2018} to search for additional unmasked spurious detections, such as unmasked bad columns. True astrophysical transients should be randomly located on each CCD, while signals due to CCD defects will be clustered in pixel coordinates. We count all detections in bins of size $8\times 16$ pixels in each $2048\times4096$ pixels CCD image, and define a Poisson distribution with the mode of transient counts in each CCD bin. We mask all bins whose counts are above the $99.99\%$ percentile of this Poisson distribution and at least twice the median bin count. Figure \ref{im:ccdmask} plots an example CCD, showing that this process finds an excess of transients in low-$x$ pixel coordinates in the CCDs, as well as bad columns that were not previously masked.

This leads to $2.7\%$ of the remaining transients being masked, and $1.2\%$ of the fakes injected in the catalog. 

\subsection{The transient catalog}
\label{sec:transientcat}
\begin{figure}[h!]
	\centering
	\includegraphics[width=0.49\textwidth]{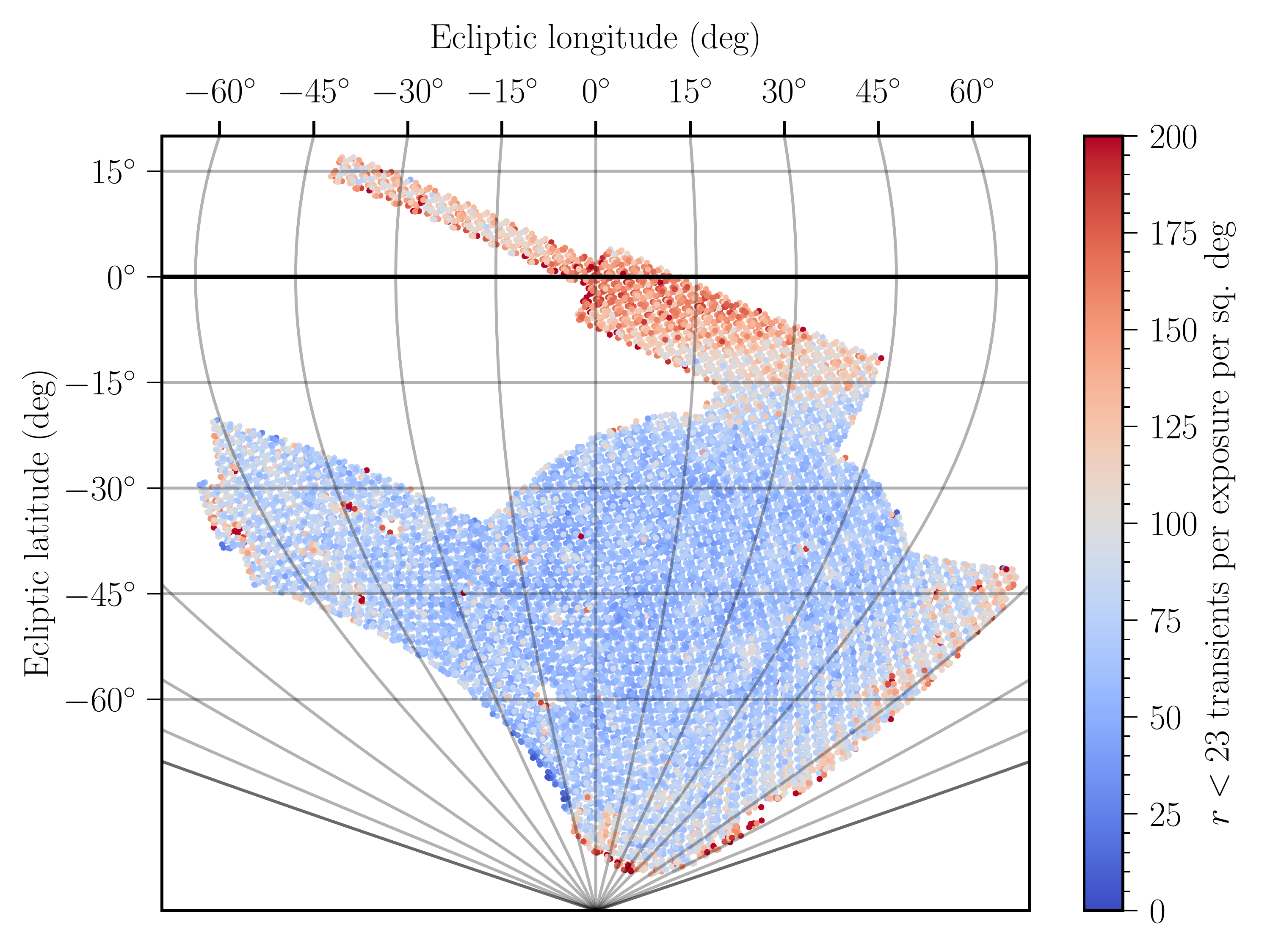}
	\includegraphics[width=0.49\textwidth]{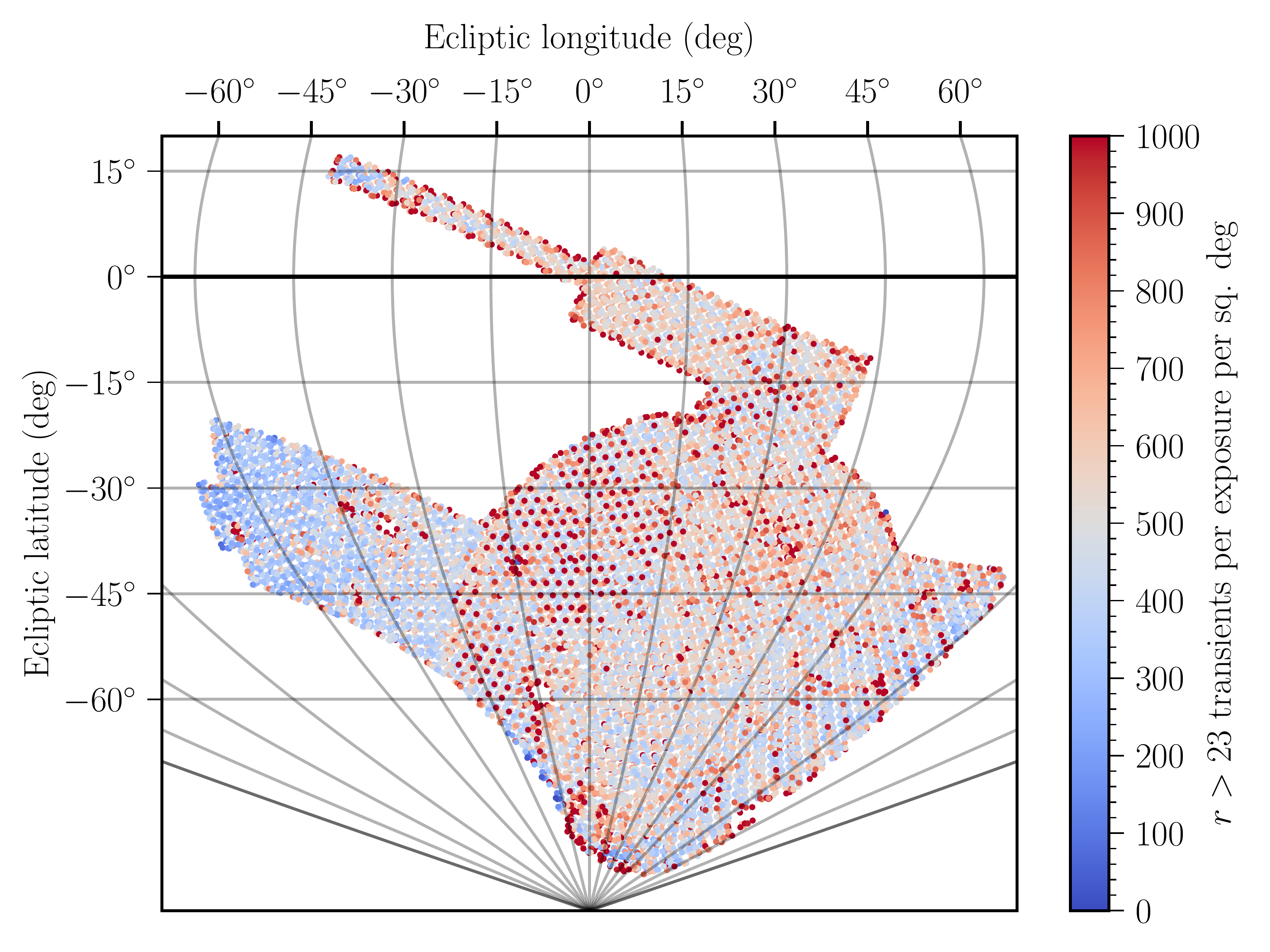}
	\caption{Sinusoidal projection of the \des\ footprint in ecliptic coordinates, each dot corresponding to one $r$ band exposure, with its color representing transient counts per exposure and per square degree. The left panel counts transients with $r < 23$~ mag. In this regime, most exposures are complete, and the transient density increases strongly toward the ecliptic plane (black line), reflecting density variations in the number of astrophysical transients (asteroids). The right panel corresponds to transients with $r > 23$. The faint end has $3-5\times$ more detections than the bright end. The lack of spatial dependence (except for a small decrease in transient counts for longitudes less than $-30\degr$) suggests that these are due to primarily to noise detections, image artifacts, and/or other non-Solar System transients.\label{im:transientmap}}
\end{figure}

The final transient catalog has 108 million sources (compare to 22 million in the Y4 catalog). Of these, 105,317 come from the fake detections injected into the SE catalog (of 110,246 originally injected), leading to an overall transient efficiency for moving object detection of 95.55\%. Figure \ref{im:transientmap} shows the distribution of transients in the survey's footprint. 

The density of bright transients shows a strong concentration toward the ecliptic plane (to $\approx 200$ per square degree for $r < 23$) at a level consistent with the expected density of asteroids \citep[\emph{e.g.}][]{Gladman2009}. The transient density increases significantly for fainter detections ($r > 23$), which dominate the catalog at all ecliptic latitudes. The mean transient density is a factor of 3--5$\times$ the transient density of the Y4 catalog, primarily a consequence of the lower detection thresholds---we have lowered the purity of true solar-system transients in an effort to increase completeness.

\subsection{Orbit linking}
\label{sec:linking}
The process of linking of detections into orbits is very similar to the one described in \citetalias{Bernardinelli2019}. Pairs of detections are found by searching in bins of inverse distance $\gamma \equiv 1/d$ \citep{Bernstein2000}, with $30 < d < 2500$ au. The detections are mapped to a frame that subtracts Earth's parallax at that distance, and so the dominant motion becomes linear in time. {The pairs are found by first finding all pairs of exposures within some time $\Delta t$ of each other that could contain a common bound TNO within the distance bin.  Then kD trees of both exposures' detections are searched for pairs of detections with separation that is}
consistent with the motion of a bound orbit. While the pair-finding algorithm remains unchanged, we have changed its implementation to use \textsc{vaex} table software \citep{Breddels2018}.  The \textsc{vaex} out-of-core functionality works well for very large datasets, e.g. some of our distance bins generate billions of pairs.

A pair of exposures strongly constrains four out of the six orbital degrees of freedom, essentially leaving distance and line-of-sight motion weakly constrained.
{The triplet stage proceeds by determining the two-dimensional region of a future exposure spanned for plausible variations of these two parameters, and locates all detections within this region.}
The algorithm of \citetalias{Bernardinelli2019} used a kD tree implementation for this stage, but for the Y6 search we devise a linear-algebra-based ``parallelogram'' search which is faster. The details are presented in Appendix \ref{sec:triplets}. We consider only triplets whose dates of observation $t_1, t_2,$ and $t_3$ satisfy
$|t_2-t_1|<60$~days and $|t_3-t_2|<60$~days.  When searching for TNOs at distances $d>50$~au, we increase these windows to 90 days, since the number of triplets decreases steeply with distance {and we can search larger time intervals without being overwhelmed by spurious triplets.}

Once a triplet of detections is found, the search then proceeds by fitting these orbits using the procedures outlined in \citet{Bernstein2000} and \citetalias{Bernardinelli2019}. An ``$n$-let'' of linked detections is fit to an orbit with the six orbital elements left free, but with a tight Gaussian prior on the inverse distance.  All exposures are then searched for transient detections lying within the $4\sigma$ predicted error ellipse of its position in that exposure. If a new transient is found to be consistent with the orbit, an $(n+1)$-let is created and the orbit is re-fit.  This is iterated until no new transients are consistent with the orbit.  Such ``terminal'' $n$-lets are retained as TNO linkage candidates.

For each linkage, we compute the following quantities:
\begin{itemize}
	\item The $\chi^2$ of the orbit fit, following the routines of \cite{Bernstein2000}. Here, the number of degrees of freedom $\nu \equiv 2\times\mathtt{NDETECT} - 6$, with $\mathtt{NDETECT}$ being the number of detections in this orbit. We reject all orbits with $\chi^2/\nu > 4$.
	\item The number of unique nights $\mathtt{NUNIQUE} \leq \mathtt{NDETECT}$ {on which detections were made.}  This is more indicative than $\mathtt{NDETECT}$ of the chances of accidental linkage of asteroid apparitions into a TNO orbit, because the \des\ observing algorithm \citep{obstac} occasionally chooses to take successive exposures (at two minute intervals) with the same pointing in a single night. The short interval between these repeated pointings means that these intra-night exposures have highly correlated chance of containing an asteroid or image defect.  We keep only orbits with  $\mathtt{NUNIQUE} \geq 7$ (this choice is discussed in Sections \ref{sec:sts} and \ref{sec:completeness}).
	\item The false-positive rate (FPR) of the linkage. With $j$ being an index over all exposures,
$A_{j,\mathrm{search}}$ being the area in that exposure consistent with the orbit fit at $4\sigma,$ and transient density $n_j$ in exposure $j$, the FPR for a spurious linkage is
	\begin{equation}
		\mathrm{FPR} = \sum_j A_{j,\mathrm{search}} n_j.
	\end{equation}
	 {We calculate the $\mathrm{FPR}$ for the detection last linked to the orbit, and retain only those with $\mathrm{FPR} < 0.02.$} A larger minimum value of $\mathrm{FPR}$ would increase substantially the number of linkages to be tested, with no significant gain in recovery rate of the synthetic objects.
	\item The $\mathtt{ARC}$, corresponding to the time between the first and last detections of the orbit, and $\mathtt{ARCCUT} < \mathtt{ARC}$, the shortest arc that remains after any single night of detections is removed from this orbit. We keep only orbits with $\mathtt{ARCCUT} > 6$~months---that is, orbits with $\ge2$ detections outside the season of the discovery triplet.
\end{itemize}

After making these cuts, all duplicate linkages are merged, and every orbit is re-fit with no priors on distance or binding energy.  This leaves {760} candidates with $\mathtt{NUNIQUE}\ge8$ and another {8321} with $\mathtt{NUNIQUE} = 7.$

\subsection{Sub-threshold significance}
\label{sec:sts}
\begin{figure}[h]
	\centering
	\includegraphics[width=0.49\textwidth]{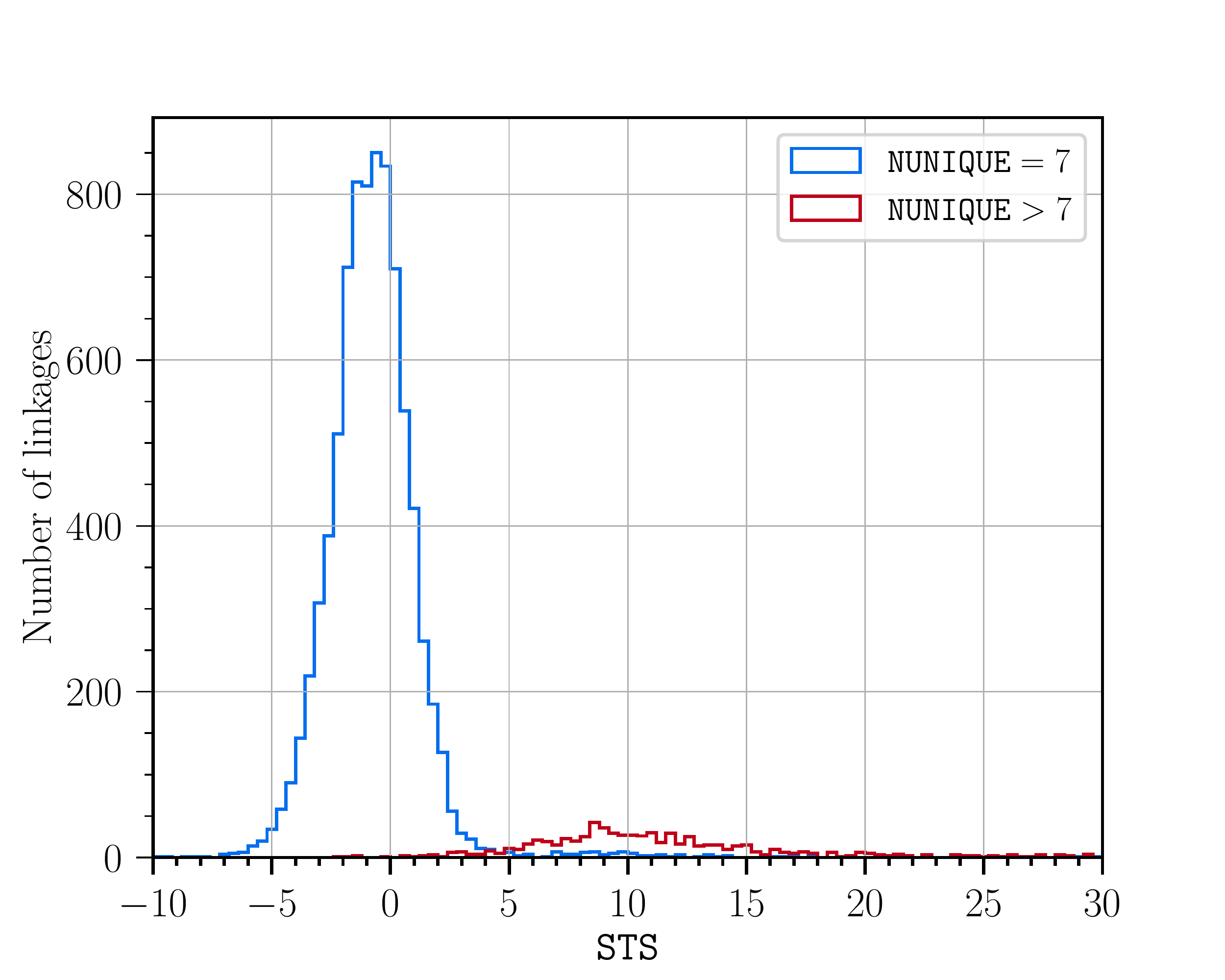}
	\includegraphics[width=0.49\textwidth]{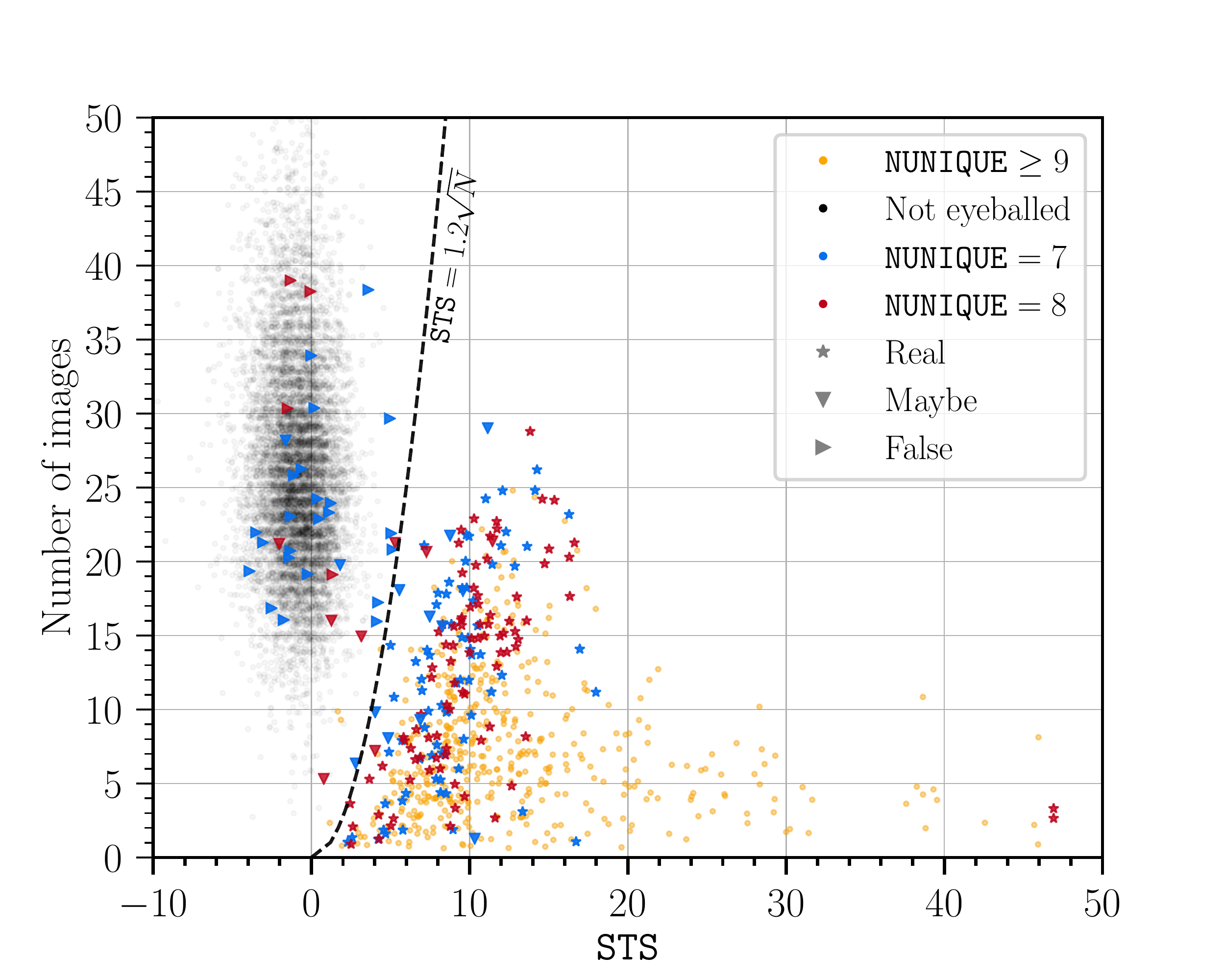}
	\caption{Results from the $\mathtt{STS}$ evaluation of 3918 sources with $\mathtt{NUNIQUE} \geq 7$. The left panel shows a histogram of the $\mathtt{STS}$ value for all sources with $\mathtt{NUNIQUE} = 7 (> 7)$ in blue (red). The right panel plots the number of images used in each $\mathtt{STS}$ stack vs the \texttt{STS} values, with color and shape encoding the \texttt{NUNIQUE} value and eyeball classification, respectively.  The $\mathtt{STS} = 1.2 \sqrt{N_\mathrm{images}}$ curve separates the bulk of low-\texttt{STS} sources from the ones considered as real in the eyeball test. The long tail of the blue histogram in the left panel and the separation of the two groupings of the right panel indicate that real sources do reliably have signal in all $griz$ images, even when they are not detected in individual exposures. The {817} objects to the right of the curve or with $\mathtt{NUNIQUE} \geq 9$ comprise our final detection catalog.\label{im:sts_y6}}
\end{figure}

To assess the reliability of a linkage, we implement the sub-threshold significance (\texttt{STS}) test from \citetalias{Bernardinelli2019}.  The $\mathtt{STS}$ is the significance of the flux peak in a stack of SE images centered on the positions predicted from the orbit fit.   The value of an object's $\mathtt{STS}$ is the signal-to-noise inside a $1\arcsec$ FWHM aperture centered in this stacked image.  The key is that all exposures which are already linked into the orbit are \emph{excluded} from the stack, as are images taken on the same night as linked detections.  This leaves only images that are statistically independent of the original linkage in terms of both shot noise and presence of asteroids or defects.  
With $N_\mathrm{images}$ $griz$ exposures kept in the $\mathtt{STS}$ stack, we examine by eye all objects with $\mathtt{STS} > 1.0 \sqrt{N_{\mathrm{images}}}$ to eliminate any images where the $\mathtt{STS}$ aperture is contaminated by static sources or unmasked artifacts.
Three of the authors (P.B., G.B., and M.S.) examined by eye postage stamps and stacked images of all candidates, and attributed to each a score of \texttt{R} (real), \texttt{M} (maybe) and \texttt{F} (false); if two of the scores agree, this is this object's final classification.  {As seen in the right-hand plot of Figure~\ref{im:sts_y6}, the curve
  $\mathtt{STS}=1.2\sqrt{N_\mathrm{images}}$ perfectly separates those graded \texttt{R} from those graded \texttt{F}.  We therefore consider as a confirmed object:
  \begin{itemize}
  \item any candidate with $\mathtt{NUNIQUE} \geq 9$ (independent of its \texttt{STS}),
  \item \textbf{and} any candidate with $\mathtt{STS} > 1.2\sqrt{N_\mathrm{images}}$ (after removing contaminated images from the \texttt{STS} stack).
  \end{itemize}
Of the 20 candidates graded \texttt{M}, 13 are above this \texttt{STS} threshold.}

Our final catalog consists of those {817} objects which pass at least one of these criteria. Figure \ref{im:sts_y6} shows the results of the $\mathtt{STS}$ of all 9081 sources with $\mathtt{NUNIQUE} \geq 7$ found in the search.

We also computed the $\mathtt{STS}$ statistic for $\approx100,000$ orbits with $\mathtt{NUNIQUE} = 6$ found at $d > 50\au$. Visual inspection of all sources with $\mathtt{STS} > 1.5 \sqrt{N_\mathrm{images}}$ yielded only one source whose STS value was not spuriously high due to contamination by unmasked artifacts or static sources.
This is contrary to the result in \citetalias{Bernardinelli2019}, where most $\mathtt{NUNIQUE} = 6$ orbits came from real linkages. Because examination of all $\mathtt{NUNIQUE} = 6$ candidates would be a large undertaking with very little gain in secure TNOs, we elect to set $\mathtt{NUNIQUE} = 7$ as our minimum threshold for the Y6 TNO search. The modest effects of this choice on TNO completeness are discussed in Section \ref{sec:completeness}.

\subsection{False negatives}
\label{sec:falseneg}
The combination of the \texttt{NUNIQUE} and \texttt{STS} criteria give us confidence that the final catalog does not contain false positive detections.  Now we ask whether we have false negatives, i.e. objects we should have detected but missed.  Our proposition is that our final catalog is almost 100\% complete for TNOs whose appearances in the transient catalog satisfy all the following criteria for being ``discoverable'':
\begin{enumerate}
\item  at least one triplet that fits the timing criteria in Section \ref{sec:linking};
\item $\mathtt{NUNIQUE}\ge 7$;
\item $\mathtt{ARCCUT}>6$~months.
\item Heliocentric distance in 2016 $>29$~au.
\end{enumerate}
\subsubsection{Synthetic TNOs}
The efficiency of our linking process can be judged, first,  by seeing whether the linker successfully found all of the TNOs whose synthetic detections injected into the SE catalogs (as per Section~\ref{sec:fakes}) created a set of transients meeting the discoverability criterion.  We find that 3710 out of 3749 ($\approx 99\%$) such objects were in fact successfully linked and ``discovered''. All of the 39 missed linkages were in regions of the footprint with high transient density, suggesting that these did not satisfy the false-positive rate criteria during the linking process.

  There is one caveat here, which is that we do not have postage stamp images for the fakes and thus cannot calculate an \texttt{STS} value.  The fairly clear separation between high- and low-\texttt{STS} populations in Figure~\ref{im:sts_y6} strongly suggests that $>99\%$ of real objects do lie above the threshold.

\subsubsection{Known objects}
We can also address the false-negative question by asking whether all previously-known TNOs that crossed the \des\ footprint made it into our catalog.  We start by comparing the Y6 catalog to the Y4 catalog.
Five objects out of the 316 that were found in the Y4 search are missed in the Y6 search. The objects, and the reason they are missing in this search, are:
\begin{itemize}
	\item 2003 QT$_{90}$ and 2013 VQ$_{46}$: these objects had $\mathtt{NUNIQUE} = 6$ in this search (as in Y4), and are on edges of the \des\ footprint, not allowing for further observations in Y6;
	\item 2010 SB$_{41}$ and (534315) 2014 SK$_{349}$, that were discovered at distances $d = 28.2 \au$ and  $d = 27.8\au$, respectively, which are too close to be found in the smallest characterized distance bin for this search;
	\item 2013 RL$_{124}$: the only possible discovery triplet for this object in Y6 included a pair of detections 73 days apart, and the maximum time difference for detections in this search for $d < 50\au$ is 60 days (see Section \ref{sec:linking}, note that this restriction was \emph{not} present in the Y4 search), meaning that triplets of this object were never searched for further detections.
\end{itemize}
We thus see that none of these five met the discoverability criterion for Y6.

We also compare our list of objects to an updated list of known objects in the Minor Planet Center using the methodology of \cite{BandaHuarca2019}. This list leads to a total of 591 objects that cross the \des\ footprint, including objects that do not satisfy our detection criteria. Of these, 356 have detection sets that are  ``discoverable'' by our criteria and {100\% of these} have indeed been retrieved in this processing.

In summary, the evidence from both simulated and real TNO observations is that our linking process is highly complete ($>99\%$) for TNOs whose brightness and geometry produce a discoverable set of entries in the transient catalog.

\section{Survey simulation}
\label{sec:surveysim}

\subsection{Simulator methods}
In order to quantify our observational biases, we have developed software that allows observations of population models to be simulated, similar to \cite{Jones2006} and \cite{Lawler2018}. Each simulated object requires a set of orbital elements, magnitudes (either apparent or absolute) in $griz$ and potential light curve, allowing for a wide range of parameters to be varied in these tests. As an example, the populations of fakes described in \citetalias{Bernardinelli2019} and Section \ref{sec:fakes} as well as the the extreme TNO ensemble of \cite{Bernardinelli2020} were realized within this framework. 

For each set of orbital elements, we find all potential observations of such a TNO, that is, we find every \des\ pointing in which such an object's position would lie inside a functional DECam CCD. We have recorded completeness estimates $\{m_{50}, c, k\}$ for all \des\ exposures used in the search, {and so we can iterate over the exposures and calculate the probability of the TNO being detected using Equation \ref{eq:probdet} given its magnitude and light curve. An additional factor of 95.5\% gives the probability that a detection will be correctly identified as a transient (as determined in Section~\ref{sec:transientcat}).  A random-number draw determines whether this TNO yields a transient detection on each of its observable exposures.}


  From the list of transient detections for a posited TNO, we determine whether the TNO is discoverable by the criteria listed in Section~\ref{sec:falseneg}.  Given the results of that discussion, we can safely assume that any posited TNO which is discoverable would, in fact, have been discovered by the survey.

{There is one caveat to our simulation algorithm, which is that the stellar objects used to determine the detection probabilities (Section~\ref{sec:det_thresh}) are, by definition, only drawn from parts of the sky that yield usable images.  In particular, regions of the sky near very bright stars or galaxies, in which neither TNOs nor static stellar sources would have been found, are not properly accounted by this method.  We are thus slightly over-estimating the parameter $c$ in Eq.~(\ref{eq:probdet}), perhaps by a few percent.  The brightest stars, galaxies, and globular-cluster regions are totally removed from \des\ processing and are not considered as ``observed''---these are properly excluded.  But the ``deserts'' around less-bright stars are not.  We expect this issue to cause, at worst, a few percent mis-estimation of our total efficiency for TNOs near the completeness limits.}

This survey simulation software, as well a tutorial for its use, will be released in \url{https://github.com/bernardinelli/DESTNOSIM}. The software works with the user stipulating a set of orbital elements or phase space coordinates, an absolute or apparent magnitude and colors for each object. Routines to include the color ranges of Equation \ref{eq:colors} and light curves are also included. The software then evaluates if this object would be detected or not, and returns this information.

\subsection{Completeness testing}
\label{sec:completeness}
We show here the results of a simulation of a large population of synthetic TNOs, and the effects of selecting distinct subsets from this sample. The population of fakes is simulated in a similar manner as in Section \ref{sec:fakes}, but this population is a factor of 100 larger (yielding $\approx 450,000$ fakes) than those injected into the SE catalogs, and covers the range $20 < m_r < 25$. This time, however, we use the detectability simulations described above rather than actually linking the detections.

\begin{figure}[h]
	\centering
	\includegraphics[width=0.49\textwidth]{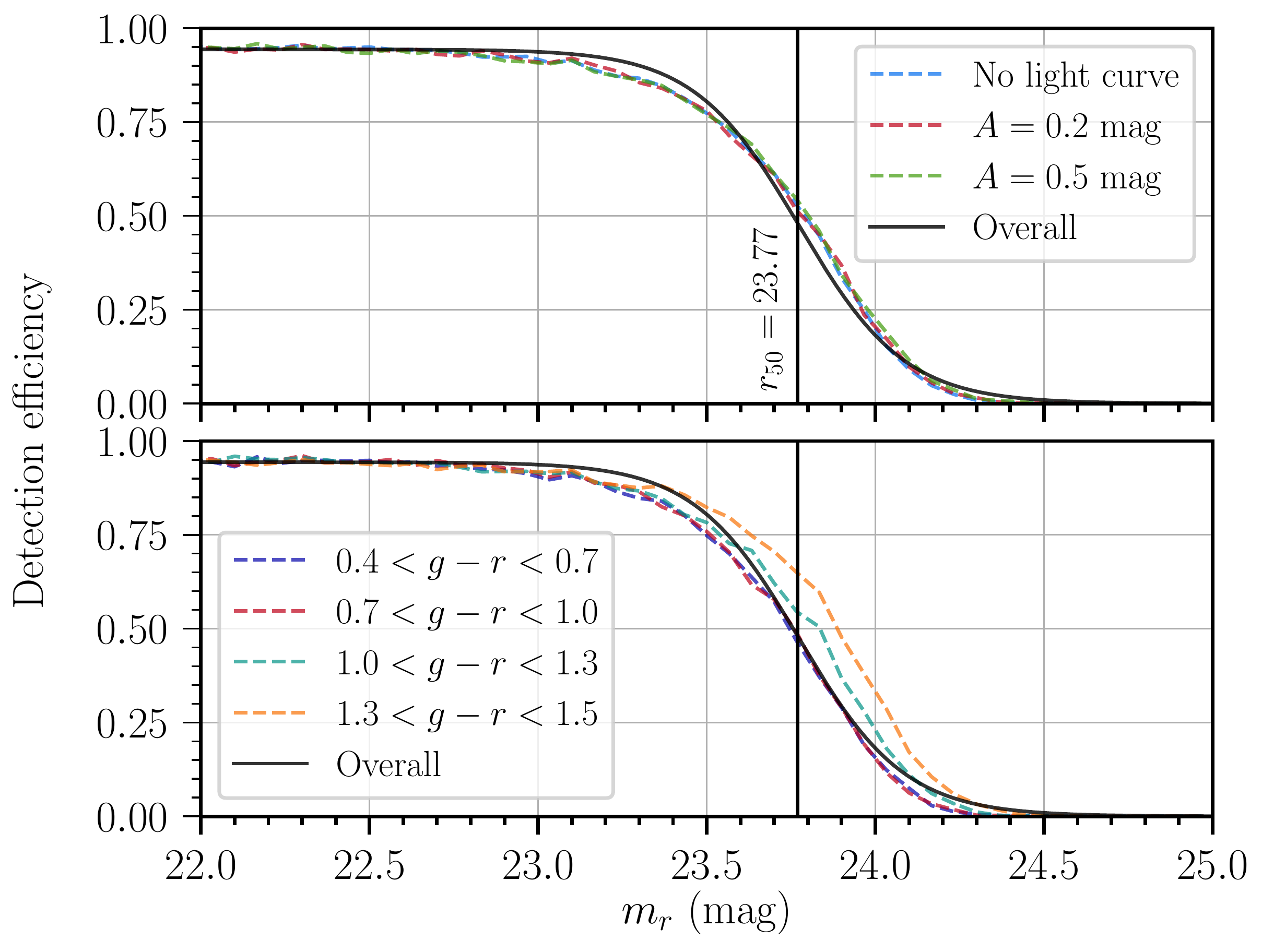}
	\includegraphics[width=0.49\textwidth]{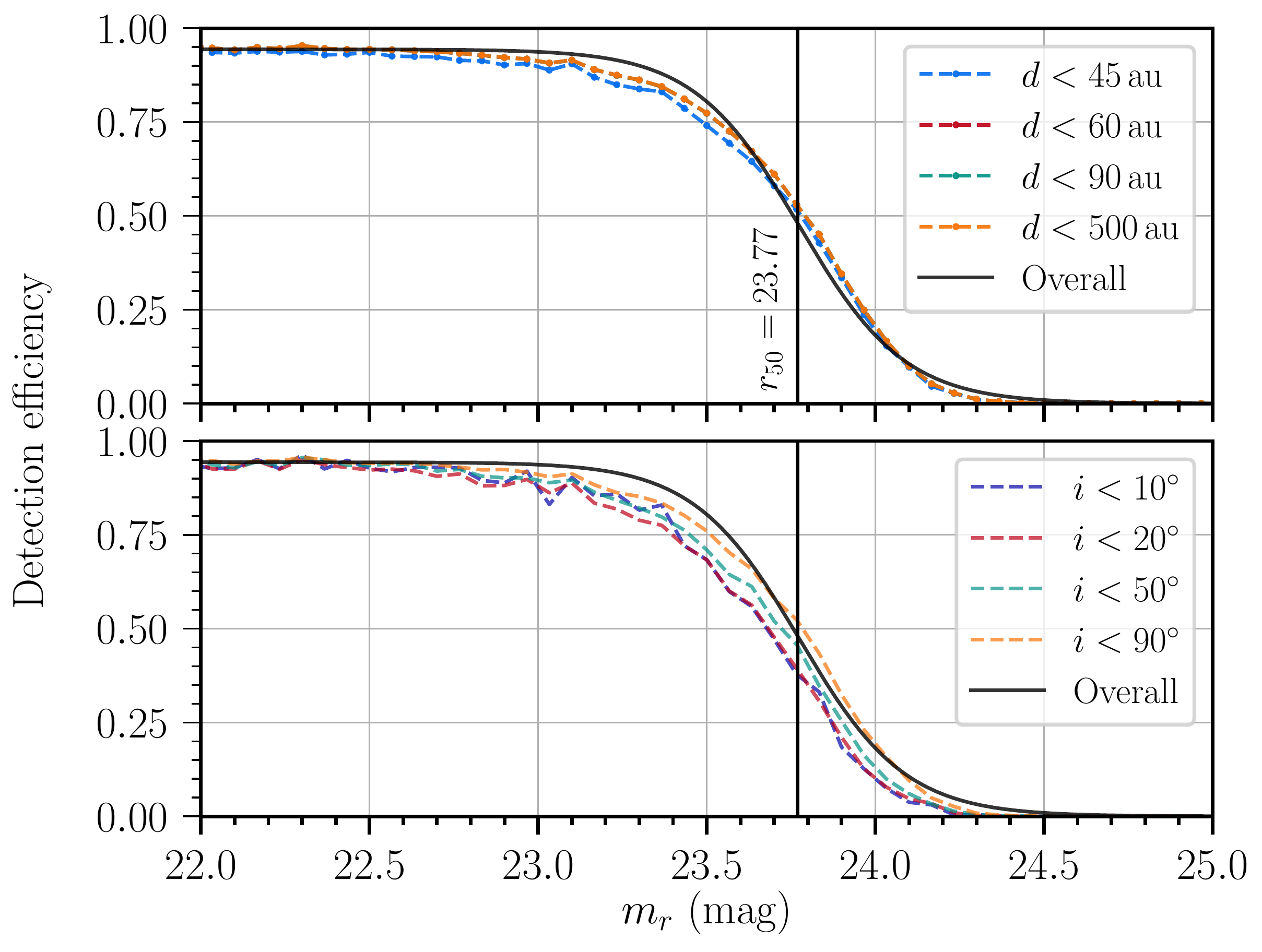}
	\caption{Measured completeness as a function of $r$ band magnitude for the same subsets of the simulated fake population presented in Table~\ref{tb:m50sim}. The left panel presents subsets of photometric properties (light curves and colors), and the right panel shows subsets of orbital properties (inclination and distance). \label{im:m50sim}}
\end{figure}

We aim to summarize the detection probability as a function of mean apparent $r$-band magnitude of each simulated source (averaging over any light curves), subdividing the simulated population into subsets of color, light-curve behavior, distance, and inclination. Figure \ref{im:m50sim} show the completeness fraction of the simulation as a function of mean $m_r$ within each subset.
We fit the logit function (Equation \ref{eq:probdet}) to the detection efficiency curves and report the fitted parameters in Table~\ref{tb:m50sim}.

It is clear from this exercise that the detection efficiency of the \des\ Y6 TNO survey is essentially a function only of mean $m_r$ for objects that do not move out of the \des\ footprint during the survey.  For the population as a whole, the survey is 50\% complete at $m_r=r_{50}=23.77$~mag.  This is a significant improvement on the value of $r_{50}=23.3$~mag in the Y4 search.

The total footprint area is $5090\deg^2$, and the value of $c=0.943$ leads to an effective search area of $4800\deg^2$.  The amplitude of a TNO light curve changes $r_{50}$ by only $\pm0.01$~mag for $0<A<0.5$~mag, and the distance to the TNO also makes no appreciable difference once the TNO is in the search region of $d>29$~au.  There is a mild dependence of $r_{50}$ on color, changing by $0.13$~mag from the bluest to reddest simulated objects.  The ``pivot point'' of the 4-band \des\ survey, i.e. the wavelength at which detection efficiency is nearly independent of TNO color, is slightly redward of the $r$ band.

The completeness drop slightly ($\approx0.1$~mag brighter) for  $i < 10,20\degr$ when compared to all the other subsets.  This is a direct effect of the shape of the \des\ footprint (Figure \ref{im:transientmap}), which is very narrow for ecliptic latitudes $|\beta| \lesssim 15\degr$: low-latitude objects (especially at closer distances) have a larger chance of moving in or out of the footprint during the \des\ duration.  This reduces the number of opportunities for the TNO to meet the detectability criteria, causing loss of fainter sources.

The inclusion of orbits with $\mathtt{NUNIQUE} = 6$ would lead to $r_{50} = 23.81$ and increase the effective search area to $4812\deg^2$, a minor gain compared to the increased burden of verification both by $\texttt{STS}$ and visual inspection of potentially hundreds of thousands of orbits. 

A very good approximation to the \des\ Y6 TNO search selection function, therefore, is that TNOs follow the logit function with $r_{50}=23.77$ mag, for any TNO that spends most of 2013--2019 inside the \des\ footprint at distance beyond 29~au.

\begin{deluxetable}{cccc||cccc}
  \tablecaption{Completeness fits for a selection of subsets of the simulated fake population. These subsets cover a different range of light curve amplitudes $A$, $g - r$ colors, inclinations $i$ and distances $d$. Other choices of parameter bins were tested (such as $e$), leading to no significant change among bins.  The $r_{50}$ entry is the light-curve-averaged $r$-band magnitude at the probability of discovery of an implanted TNO is 50\%, $c$ a scaling factor for the completeness curve, and $k$ the transition sharpness (see Equation \ref{eq:probdet}).
    \label{tb:m50sim}}
    	\tabletypesize{\footnotesize}
	\tablehead{\colhead{Subset of the simulation} & \colhead{$r_{50}$} &
        \colhead{$c$} & \colhead{$k$} & \colhead{Subset of the simulation} & \colhead{$r_{50}$} &
        \colhead{$c$} & \colhead{$k$}}
	\startdata
	\textbf{All fakes} & \textbf{23.77} & \textbf{0.943} & \textbf{6.37} & $ i < 10\degr $ & 23.67 & 0.944 & 6.54\\
	No light curve & 23.76 & 0.944 & 6.52 & $ i < 20 \degr$ & 23.68 & 0.938 & 6.24 \\
	$A = 0.2 \, \mathrm{mag}$ & 23.77 & 0.943 & 6.38 & $ i < 50 \degr $ & 23.72 & 0.946 & 6.28 \\
	$A = 0.5 \, \mathrm{mag}$ & 23.78 & 0.944 & 6.24 & $ i < 90 \degr $ & 23.77 & 0.951 & 6.43 \\
	$0.4 < g- r < 0.7$ & 23.75 & 0.950 & 6.73 & $ d < 45 \au$ & 23.76 & 0.930 & 6.21 \\
	$0.7 < g - r < 1.0$ & 23.75 & 0.952 & 6.80 & $ d < 60 \au$ & 23.77 & 0.943 & 6.37 \\
	$1.0 < g - r < 1.3$ & 23.81 & 0.951 & 6.70 & $ d < 90 \au$ & 23.77 & 0.943 & 6.37 \\
	$ 1.3 < g - r < 1.5$ & 23.88 & 0.948 & 6.92 & 	$ d < 500 \au$ & 23.77 & 0.943 & 6.37 
	\enddata
\end{deluxetable}

\section{Catalog of \des\ TNOs}
\label{sec:catalog}

\subsection{Dynamical classification}
\label{sec:dynclassy6}

We divide the catalog of confirmed Y6 TNOs into dynamical classes following a methodology similar to \cite{Gladman2008} and \cite{Khain2020}. We sample the barycentric Cartesian phase space position and covariance matrix of the orbit solution at its solution epoch ($t_0$) to generate 30 clones of each orbit, and integrate these for 10 Myr in time steps of 30 days using the \textsc{WHFast} implementation of the sympletic Wisdom-Holman mapping \citep{Wisdom1991,Rein2015}, part of the \textsc{Rebound} $N$-body simulator \citep{rebound}. The simulation is such that Jupiter, Saturn, Uranus and Neptune are treated as active particles and the mass of the terrestrial planets is added to the solar mass, as in the orbit solutions \citep{Bernstein2000}. 

Similarly to \cite{Elliot2005}, for each clone, we check all resonant arguments of the form
\begin{equation}
	\sigma = p \lambda - q \lambda_\mathrm{N} + m \varpi - n \varpi_\mathrm{N} + r \Omega - s \Omega_\mathrm{N}, \label{eq:res}
\end{equation}
where $\varpi \equiv \Omega + \omega$ is the longitude of perihelion and $\lambda \equiv \varpi + \mathcal{M}$ the mean longitude. The subscript $\mathrm{N}$ refers to Neptune's orbital elements, and we check all  $p,q,|m|,|n|,|r|\in [1,29]$ such that $p - q + m - n + r = 0$. For convenience, we define $s = 0$, as the number of possible resonant arguments becomes significantly larger if this parameter is also left free. 

We use an automated resonance identification algorithm similar to the one outlined in \cite{Khain2020}: for each resonance whose nominal orbital period $P_\mathrm{res}$ are $|P_\mathrm{res} - P(t_0)| < 0.15 P_\mathrm{N}$, that is, whose orbital periods are up to 15\% of Neptune's orbital period away from the resonance, we construct a two-dimensional histogram of $(t,\sigma)$, and, for each time slice, we check whether the values of $\sigma$ are bounded, that is, we check if there are multiple empty $\sigma$ bins in each $t$ slice. To achieve this, the bins require 1000 points in each window of 200 kyr, and we manage to identify resonances independently of the libration center. If the largest contiguous interval {of libration} spans more than 90\% of the full integration, we consider this clone to be resonant. 

We classify the non-resonant objects as follows: 
\begin{itemize}
\item Objects with $a(t_0) > 2000 \au$ are classified as Oort cloud objects;
\item Objects with $a(t_0) < a_\mathrm{N}$ are considered Centaurs;
\item As in \cite{Khain2020}, an object is considered to be scattering if it experiences excursions in semi-major axis as a function of time $a(t)$ such that
\begin{equation}
	\frac{\max{|{a(t) - a(t_0)}|}}{a(t_0)} > 0.0375.
\end{equation}
\item Following \cite{Gladman2008} and \cite{Khain2020}, detached objects are non-scattering, non-resonant orbits that have $e(t_0) > 0.24$, 
\item Objects that do not fall into these categories are classified as Classical.
\end{itemize}

Each object gets assigned the dynamical class indicated by the behavior of the majority of its clones. As in \cite{Khain2020}, we require that at least 50\% of the clones to present scattering behavior to assign this classification, and resonant objects are called ``candidates'' if over 50\% of its clones exhibit resonant behavior, while those with over 80\% of the clones showing libration of the same resonant argument are identified as securely resonant. A minority of objects do not satisfy either dynamical requirement for the majority of its clones, and receive an ``insecure'' classification. 

\subsection{Sample of TNOs}

The final catalog contains 817 objects. Of these, 506 have not been found in the Y4 processing, including the 461 newly identified objects. {One of the discoveries is a Centaur, and one is an Oort-cloud comet discovered interior to the $d>29$~au region in which we did a complete search for bound sources.  Therefore there are {815} TNOs in the fully-characterized sample.
This is the second largest catalog of TNOs to date \citep[OSSOS,][has 840 objects, {818} in the characterized TNO phase space]{Bannister2018}, and the largest with multi-band photometry accompanied by orbital arcs of multiple years. Figure \ref{im:aei_y6} shows the semi-major axes $a$, eccentricities $e$ and inclinations $i$ for all bound objects, as well as their dynamical classifications. This figure excludes the one object with $a>10,000\au$, discussed below. Table \ref{tb:objects} describes the parameters included in the released catalog for each object, and the full table is provided in machine-readable format. A summary of the dynamical classification is presented in Table \ref{tb:pop}.
\begin{figure}[ht!]
	\centering
	\includegraphics[width=1\textwidth]{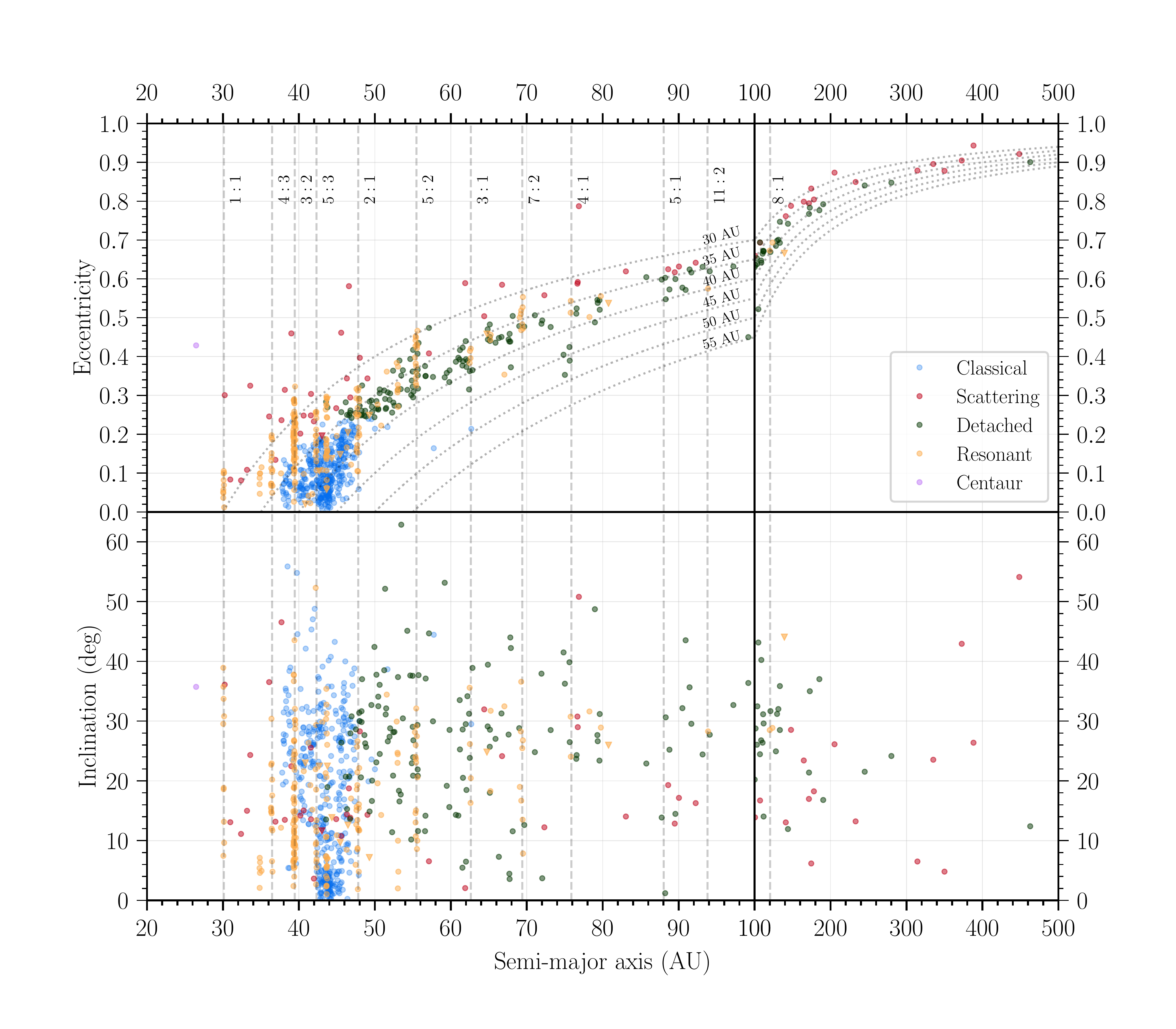}
	\caption{Semi-major axes, eccentricities and inclinations of the 816 TNOs (excluding the object found in a cometary orbit) found in this search, color-coded by dynamical class. The dotted lines represent constant perihelion $q = a(1-e)$, and the vertical dashed lined the approximate locations of some $p$:$q$ mean motion resonances. Since many of these have similar $a$, we do not indicate all occupied resonances. The solid circles represent objects with secure dynamical classifications, and the triangles objects whose classification is insecure (including resonant candidates). Table \ref{tb:pop} presents the number of objects found in each dynamical class.\label{im:aei_y6}}
\end{figure}

\begin{deluxetable}{ccc}
  \tablecaption{\des\ trans-Neptunian objects for the Y6 release. The description of each column is given here.
\textbf{The Table is provided in FITS format as an ancillary file or in the journal version of the paper.}  
     All of the elements reported are barycentric and refer to epoch 2016.0, and their uncertainties correspond to the $1\sigma$ uncertainty marginalized over other orbital parameters. \label{tb:objects}}
	\tablehead{\colhead{Column name}  & \colhead{Unit} & \colhead{Description}}
	\startdata
	\texttt{MPC} &  & Minor Planet Center object designation \\
	$a$ (\texttt{a}) &  au & Semi-major axis of the best-fit orbit \\
	$\sigma_a$ (\texttt{sigma\_a}) &  au & Uncertainty in $a$ \\
	$e$ (\texttt{e}) & & Eccentricity \\
	$\sigma_e$ (\texttt{sigma\_e})& & Uncertainty in $e$ \\
	$i$ (\texttt{i}) & $\deg$ & Inclination \\
	$\sigma_i$ (\texttt{sigma\_i}) & $\deg$ & Uncertainty in $i$ \\ 
	$\omega$ (\texttt{aop}) & $\deg$ & Argument of perihelion \\
	$\sigma_\omega$ (\texttt{sigma\_aop}) & $\deg$ & Uncertainty in $\omega$ \\
	$\Omega$ (\texttt{lan}) & $\deg$ & Longitude of ascending node \\
	$\sigma_\Omega$ (\texttt{sigma\_lan}) & $\deg$ & Uncertainty in $\Omega$ \\
	$T_p$ (\texttt{T\_p}) & UTC Modified Julian date & Time of perihelion passage  \\ 
	$\sigma_T$ (\texttt{sigma\_T}) & days & Uncertainty in $T_p$\\
	$q$ (\texttt{q})& au & Perihelion distance \\
	$\sigma_q$ (\texttt{sigma\_q}) & au & Uncertainty in $q$ \\
	$d$ (\texttt{d}) & au & Discovery distance (geocentric) \\ 
	$\sigma_d$ (\texttt{sigma\_d}) & au & Uncertainty in $d$ \\
	$\Delta$ (\texttt{delta}) & au & Discovery distance (heliocentric) \\
	$\sigma_\Delta$ (\texttt{sigma\_delta}) & au & Uncertainty in $\Delta$ \\
	$m_r$ (\texttt{m\_r}) & mag & Mean $r$ band magnitude \\
	$\sigma_m$ (\texttt{sigma\_m}) & mag & Uncertainty in $m_r$ \\
	$H_r$ (\texttt{H\_r}) & mag & Absolute magnitude in band $r$ \\
	$\sigma_H$ (\texttt{sigma\_H}) & mag & Uncertainty in $H_r$ \\
	\texttt{NUNIQUE}  & & Number of unique nights of detections \\
	\texttt{NDETECT} & & Number of detections \\ 
	$\chi^2$ (\texttt{CHI2}) & & $\chi^2$ of the orbit fit, where $\nu = 2\times \mathtt{NDETECT} - 6$ \\ 
	$x,y,z$ (\texttt{x}, \texttt{y}, \texttt{z}; 3 columns) & au & ICRS-oriented positions\\
	$v_x, v_y, v_z$ (\texttt{v\_x}, \texttt{v\_y}, \texttt{v\_z}; 3 columns) &  $\text{au}/\text{year}$ & ICRS-oriented velocities \\
 	$\Sigma_{\mu,\nu}$ (\texttt{Sigma\_mu\_nu}; 21 columns) & $(\text{au},\text{au}/\text{yr})^2$ & $\mu,\nu$ element of the state vector covariance matrix.\\
	\texttt{Class} & & Dynamical classification  \\
	\texttt{Notes} & & Notes on the object$^{1}$
	\enddata 
	\tablenotetext{1}{Insecure dynamical classifications include \texttt{ins} in this column, non-characterized objects have \texttt{nc}, and resonant objects also have their $\{p,q,m,n,r,s\}$ resonant arguments listed.}
\end{deluxetable}

\begin{deluxetable}{cc|cc}[h!]
	\tablecaption{Number of objects per dynamical classification for the 817 objects. The resonant objects are presented in order of increasing semi-major axis, with the approximate value presented in parenthesis. \label{tb:pop}}
	\tablehead{\colhead{Dynamical class} & \colhead{Number of Objects} & \colhead{Dynamical class} & \colhead{Number of Objects} }
	\startdata
	Classical belt & 382 ($+1$ insecure) & Scattering & 51 ($+1$ insecure) \\
	Detached & 155  & Centaur & 1  \\
	Oort cloud & 1 \\
	\hline \hline
	Mean-motion resonance & Number of objects & Mean-motion resonance & Number of objects \\ \hline
	1:1 (30.1 au) & 10 & 13:6 (50.9 au) & 1 \\
	5:4 (34.9 au)  & 6 & 11:5 (50.4 au) & 1\\
	4:3 (36.3 au) & 14 & 9:4 (51.7 au) & 1  \\
	7:5 (37.7 au) & 1 & 7:3 (52.9 au) & 7 \\
	3:2 (39.4 au) & 69 & 5:2 (55.4 au) & 15 \\
	26:17 (39.9 au) & 1 & 	3:1 (62.6 au) & 4 \\
	23:15 (40.0 au) & 1 & 	19:6 (64.9 au) & 1 candidate \\
	11:7 (40.7 au) & 1 & 	16:5 (65.4 au) & 2 \\
	19:12 (40.9 au) & 1 candidate & 	10:3 (67.1 au) & 1 \\
	5:3 (42.3 au) & 14 + 1 candidate & 	7:2 (69.4 au) & 8\\
	22:13 (42.7 au) & 1 candidate &	4:1 (75.8 au) & 2 \\
	12:7 (43.1 au) & 1  & 	21:5 (78.4 au) & 1 \\
	7:4 (43.7 au) & 25 + 2 candidates & 	13:3 (80.0 au) & 1 \\
	29:16 (44.8 au) & 1 & 	22:5 (80.8 au) & 1 candidate \\
	11:6 (45.1) au) & 1 + 1 candidate & 11:2 (93.8 au) & 1 \\
	13:7 (45.5 au) & 1 candidate &	8:1 (120.4 au) & 1 \\
	2:1 (47.7 au) & 21 & 25:3 (123.7 au) & 1 \\
	23:11 (49.2 au) & 1 & 10:1 (139.7 au) & 1 candidate \\
	21:10 (49.3 au) & 1 candidate \\
	\hline \hline 
	Total & 817\\
	\enddata
\end{deluxetable}

The following properties of the sample, as well as particularly interesting objects, are highlighted: 
\begin{itemize}
	\item {Object C/2014 UN$_{271}$ (Bernardinelli-Bernstein)}, inbound on a near-parabolic cometary orbit, with $e = 0.999419 \pm 0.000011$, $q = 10.95\au$, $i = 95.56\degr$, and nominal discovery distance $d = 27.53\au$, incoming from the Oort cloud. This is also the only one of the objects discovered at $d<29$~au, and is therefore not part of our well-characterized search space.
	\item Two new discoveries at $d > 70 \au$: the first is our largest-$q$ object with $q = 54.31\au$, discovered at a distance of $79\au$ (the third most distant object from this search). While most clones of this orbit are non-resonant and this object is classified as detached, a portion of its clones were identified to be in the 6:1 resonance. The second is a scattering object with low-$q$ discovered far from its perihelion, at $d=73\au$.
	\item Several new objects that securely occupy distant, high-order mean motion resonances: for example, we report one object in each of the 8:1 and 25:3 mean-motion resonances (see \citealt{Volk2018} for a discussion of a sample of 9:1 resonators). An object is also classified as a resonant candidate for the 10:1 resonance (that is, over half of its clones are in the 10:1 resonance, but not more than 80\% as defined above).
	\item 9 new ``extreme'' TNOs (eTNOs, $a > 150$, $q>30\au$), increasing the total measured by \des\ to 16. Of these, 9 (4 newly reported here) also have $a>230\au$ and are of interest to the Planet 9 hypothesis \citep[][see also \citealt{Shankman2017}, \citealt{Bernardinelli2020} and \citealt{Napier2021}]{Batygin2019}. These objects are discussed in Section \ref{sec:etno}.
	\item Four new Neptune Trojans, increasing our sample size to ten, as well as the unrecovered 2013 RL$_{124}$.
	\item Several new objects with high inclinations, including a detached object with $a = 53.28\au$ and $i=62.84\degr$, the most inclined object with $q>30\au$ to date, and two new objects in classical orbits with inclination $i > 54 \degr$.
	\item Two objects in low-$e$ (classical) orbits with $a > 55\au$, on orbits with high inclinations ($i > 29\degr$) and perihelia ($q > 48\au$), being similar to the detached population, despite their assigned dynamical classes.
	\item A total of 155 objects securely classified as detached, the largest sample of this population found by single survey to date.
\end{itemize}

The discovery distance and magnitude of these objects are shown in Figure \ref{im:y6_mags}, as well as the 50\% completeness limits of this search and the Y4 search. Most objects new to this search are found at $m_r > r_{50,\mathrm{Y}4} \approx 23.3$, as expected. However, 30 new ``bright''($m_r < 23$) objects were found in the Y6 search, since the two additional years of data allow for new objects to enter the survey's footprint for long enough to meet our selection criteria (see Figure \ref{im:y6_mags}).

\begin{figure}[h!]
	\centering
	\includegraphics[width=0.49\textwidth]{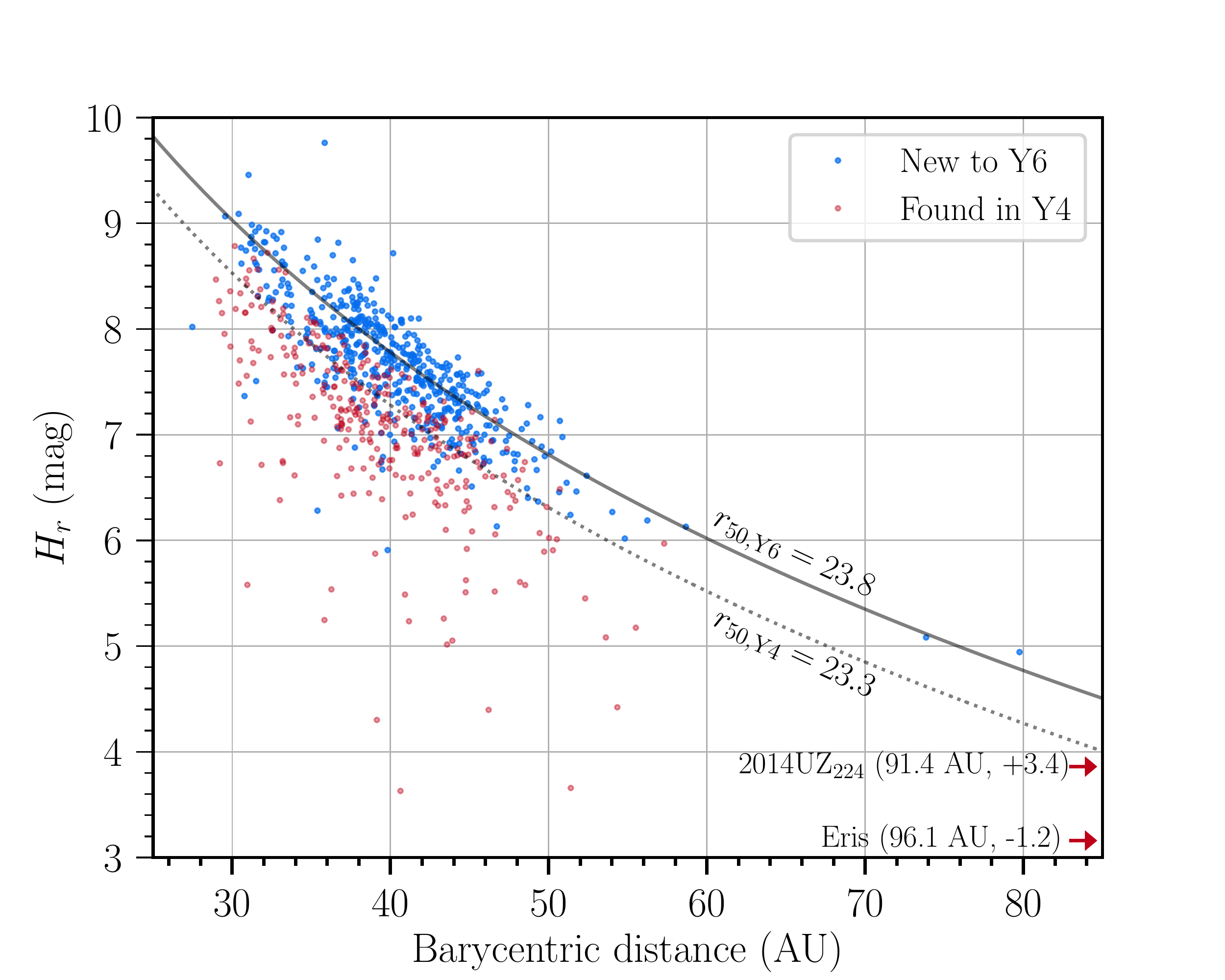}
	\includegraphics[width=0.49\textwidth]{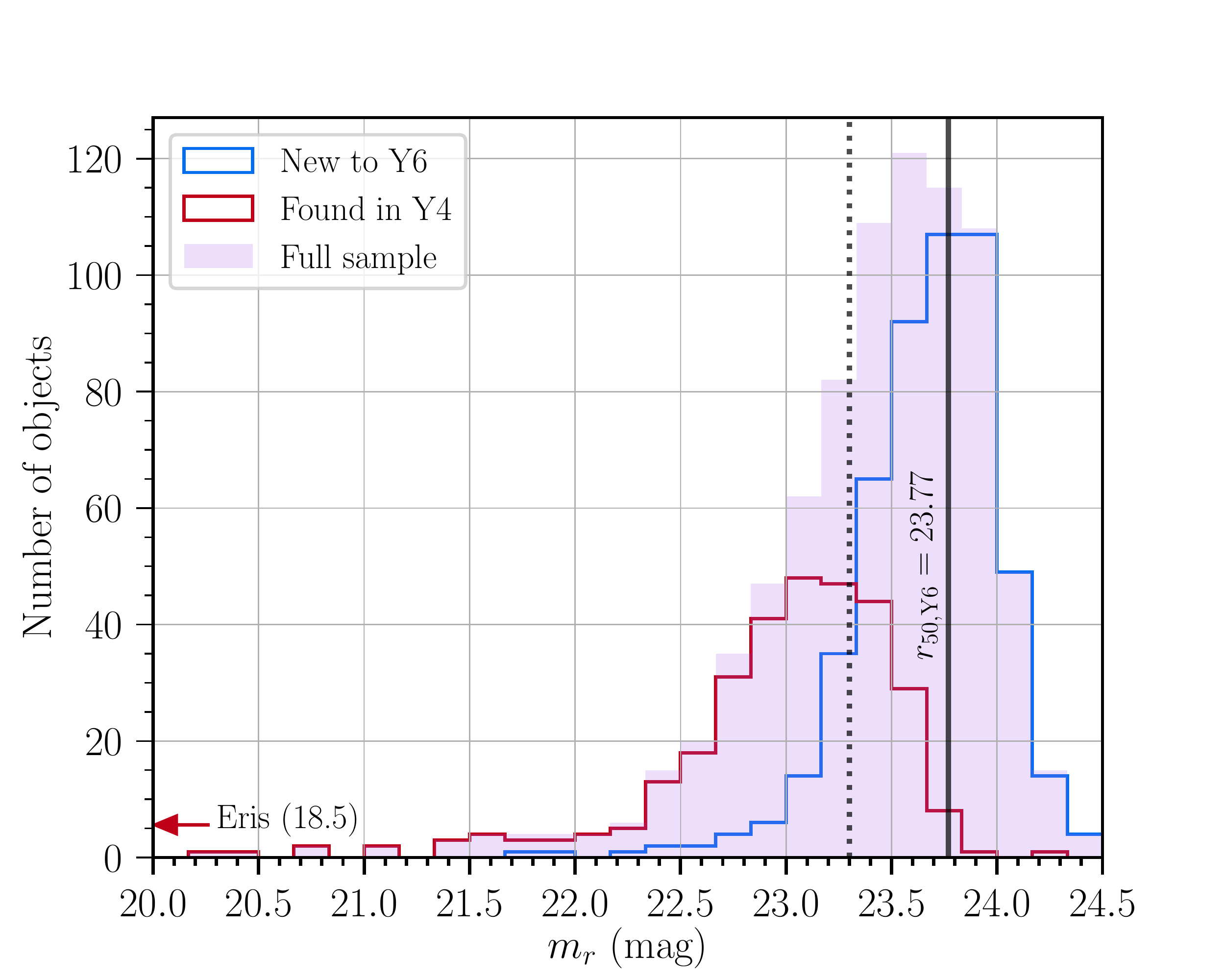}
	\caption{\emph{Left:} Absolute magnitude $H_r$ versus barycentric discovery distance for the 817 objects reported here. The blue dots correspond to objects new to this search, while the red correspond to those that were present in the Y4 catalog. The dotted curve shows the $r_{50} = 23.3$ of the Y4 search, and the solid black curve the $r_{50} = 23.77$ of this search. The absolute magnitudes are found by taking the mean flux in each exposure corrected to a nominal geocentric distance $d$ and heliocentric distance $\Delta$. \emph{Right:} Histogram of apparent magnitudes $m_r$ for the objects found in this search. The solid purple histogram shows the full sample, while the lined histograms show the sample found in the Y4 search and the sample new to this Y6 search. The black lines (dotted and solid) are the same as in the left panel. While the majority of the new objects were discovered at $m_r > r_{50,\mathrm{Y}4}$, 56 were found at magnitudes brighter than $r_{50,\mathrm{Y}4}$.
\label{im:y6_mags}}
\end{figure}

\section{Initial implications}
\label{sec:implications}
\subsection{Comparison to CFEPS model}

The most accurate model to date for the classical TNOs is that derived by the CFEPS project \citep{Kavelaars2009,Petit2011,Petit2017,Gladman2012}. In their ``L7'' model,  the classical population is described as a combination of a low-$e$, low-$i$ ``kernel'' in a narrow $a$ range, a ``stirred'' component of stable, low-$i$ orbits and a high-$i$ ``hot'' component.  The CFEPS-L7 model is found to be consistent with the classical TNOs detected in the OSSOS survey by \citet{Bannister2016}.  Here we ask whether the main belt classical TNOs ($40 < a < 47\au$, roughly the region between the 3:2 and 2:1 resonances) in the \des\ Y6 catalog are consistent with the CFEPS-L7 model, or if these improved data require the model to be revised. 

The CFEPS-L7 model consists of a simulated population of 26,031 members of the main belt with $H_g<8.5$.  We simulate the \des\ observation of this proposed population as in Section \ref{sec:surveysim}. The orbital elements and $H_g$ provided for each body by CFEPS-L7 need to be assigned colors for the \des\ simulation.  For each member of the kernel and stirred components, we draw a synthetic $g-r$ color from the observed distribution in the \des\ classicals with $i<5\degr$, and for the hot component we draw colors from the \des\ classicals at $i>5\degr.$  The $g-i$ and $g-z$ colors are obtained following in Equation \ref{eq:colors}. The simulation predicts 695 \des\ detections, about $3\%$ of the total CFEPS-L7 population in this region.

We compare this simulated \des\ classical TNO population to the 252 real \des\ Y6 objects in the same $a$ range with $H_g<8.5.$  The tests presented here only use the normalized distributions in each parameter, so the difference in total number of objects is not considered in the tests. Figure \ref{im:cfeps} shows the cumulative distributions in $a$, $e$, $i$, and $H_g$ expected from the L7 model in comparison to the observed \des\ detections. A Kolmogorov-Smirnov test \citep[KS test,\emph{e.g.}][]{Press2007} leads to $p$-values $ = \{0.0257, 0.0382, 0.0046, <10^{-5} \}$ for the $\{a,e,i,H_g\}$ distributions, respectively.  The observed $e, i,$ and $H_g$ distributions are incompatible with the CFEPS-L7 model, which  is not surprising considering the large increase in information on high-inclination TNOs in the \des\ data compared to the CFEPS data to which the L7 model was initially fit.  

We investigate whether a simple reweighting of the L7 sub-populations can improve agreement with the \des\ obserations. We reparametrize the L7 classical model by changing the fraction $f_\mu$ of the stirred, hot, and kernel components $\mu$ as 
\begin{equation}
	\frac{\mathrm{d}n}{\mathrm{d}i} = f_\mathrm{s} \frac{\mathrm{d}n_\mathrm{s}}{\mathrm{d}i} + f_\mathrm{h} \frac{\mathrm{d}n_\mathrm{h}}{\mathrm{d}i} + f_\mathrm{k} \frac{\mathrm{d}n_\mathrm{k}}{\mathrm{d}i}
\end{equation}
with the constraint that $f_\mathrm{s} + f_\mathrm{h} + f_\mathrm{k} = 1$. A least-squares fit of $\mathrm{d}n/\mathrm{d}i$ to the measured inclination distribution leads to  $f_\mathrm{h} = 0.653$, $f_\mathrm{s} = 0.222$ and $f_\mathrm{k} =0.125$, 
compared to $f_\mathrm{h} = 0.51$, $f_\mathrm{s} = 0.38$ and $f_\mathrm{k} = 0.11$ in the original CFEPS model.\footnote{Note that the stirred and kernel components share the same inclination distribution in the model, and so any difference in $\mathrm{d}n/\mathrm{d}i$ comes from the selection functions for each component.} Repeating the KS test with these new fractions leads to KS test $p$-values $ = \{0.0018, 0.7792, 0.0134, \leq 10^{-5}\}$. The model reproduces the measured $e$ distribution satisfactorily, but rejects the new $a$, $i$ and $H_g$ distributions. No simple subpopulation reweighting can bring the L7 model into agreement with the observations. The \des\ Y6 catalog, then, will enable considerable refinement of TNO population models, particularly in combination with the OSSOS data.

Repeating these tests with shallower $H_g<8.0$ samples (351 simulated \des\ detections, 131 real ones)---this being the limit of the CFEPS data to which the model was fit---yields consistency between the data and the model. The $p$-values are $\{0.079, 0.338, 0.308, 0.067\}$ in ${a,e,i,H_g}.$

\begin{figure}[h]
	\centering
	\includegraphics[width=\textwidth]{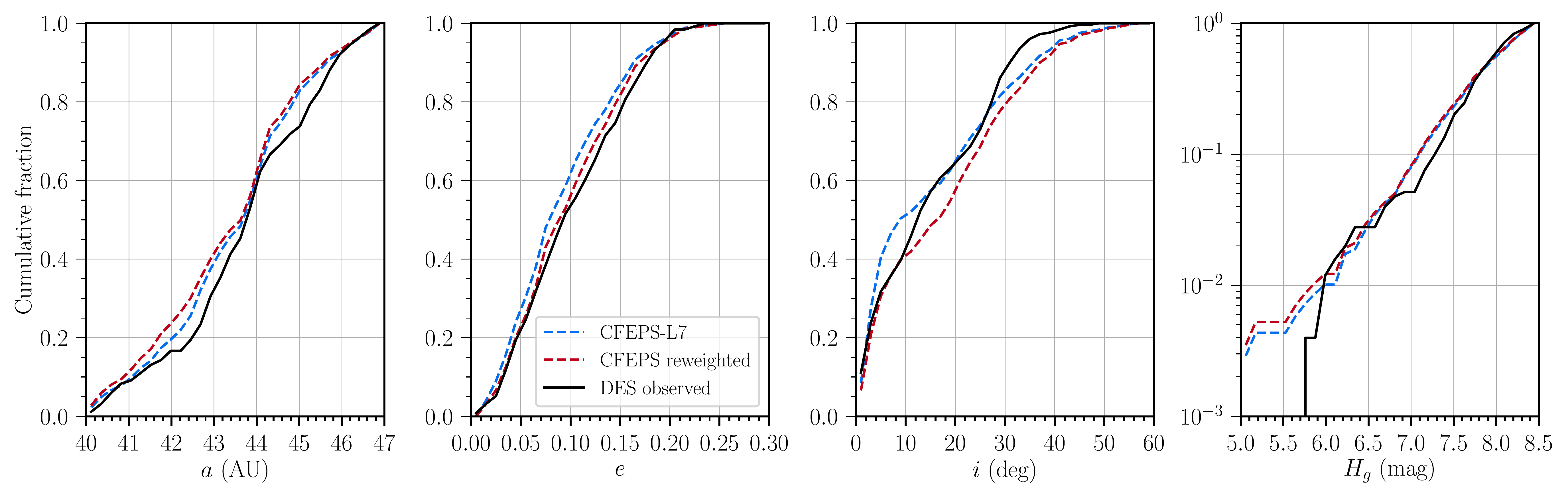}

	\caption{Comparison between the CFEPS-L7 classical objects with $40\au<a<47\au$ simulated into the \des\ Y6 and the \des-detected classical TNOs in this range. Each panel presents the cumulative distribution in $\{a,e,i,H_g\}$, with the blue lines representing the baseline model and the red lines representing the reweighted models. The black line corresponds to the \des\ observations.\label{im:cfeps}}
\end{figure}

\subsection{Isotropy of extreme TNOs revisited}
\label{sec:etno}
We repeat the test of eTNOs' isotropy presented in \cite{Bernardinelli2020}, which asked whether the population of TNOs with $a>150\au$ and $q> 30\au$ shows deviation from an underlying uniform distribution in $\omega$, $\Omega,$ and $\varpi \equiv \Omega + \omega.$ Deviations from such isotropy serve as the primary motivation for the Planet 9 hypothesis \citep{Batygin2016,Batygin2019}. The underlying population model chosen for these objects is such that we clone the detected $\{a_j,e_j,i_j,H_j\}$ for each object $j$, and randomize their $\omega$, $\Omega,$ and mean anomaly $\mathcal{M}$. \des\ Y6 observations of these objects are then simulated as in Section \ref{sec:surveysim}, and we compute the probabilities $p(\theta|s)$ of detecting an object with angle $\theta \in \{\omega,\Omega,\mathcal{M}\}$ conditioned on a successful detection $s$. We compare $p(\theta|s)$ to the empirical distributions coming from the true detected eTNOs using Kuiper's variant of the KS test \citep{Kuiper1960}, as described in \cite{Bernardinelli2020}. 

We present here the results for the population with $a>150\au$ and $q>30\au$ (16 objects), as in \cite{Shankman2017} and Case 1 of \cite{Bernardinelli2020}, as well as a more restrictive $a>230\au$ selection (9 objects), as in \cite{Brown2019} and \cite{Napier2021}, where the clustering signal is supposed to be stronger\footnote{We note that \cite{Brown2021} present a similar test on objects with $a>150\au$ and $q>42\au$.}. We obtain, for $\{ \omega, \Omega, \varpi\}$, $p$-values $= \{ 0.9224, 0.9514, 0.8788 \}$ for the $a>150\au$ sample, and $p$-values $= \{0.1355, 0.0144, 0.5851 \}$ for the $a>230\au$ sample. The $a>150\au$ sample is consistent with an underlying isotropic distribution, that is, the null hypothesis that the observed values come from the parent isotropic distribution cannot be rejected. The $a>230\au$ case shows poorer agreement with isotropy in $\Omega$, {but we note first that there is a $1 - (1-0.0144)^6 = 8\%$ chance that one of our 6 $p$-values would be this low for truly isotropic data (this probability becomes $4\%$ considering only the 3 $a>230\au$ tests); and second,} the isotropic hypothesis is fully acceptable in $\varpi$, the variable in which the clustering signal is supposed to be the strongest \citep[see discussion in][]{Bernardinelli2020} and $\omega$. We repeat the $f$-test of \cite{Bernardinelli2020}, where the total likelihood of detection for all objects is compared to the likelihood of an ensemble of clones drawn from each object's selection function. We have that $f(\Omega) = 0.287$, that is, $28.7\%$ of the sets of clones yield a lower likelihood than one measured for the detected objects. This test indicates that the {Kuiper statistics are not being driven by individual highly unlikely points.   This result, then, remains in agreement with \cite{Shankman2017}, \cite{Bernardinelli2020} and \cite{Napier2021}: the apparent clustering in orbital element space of these objects is consistent with the selection functions of the surveys and does not conclusively demand a massive perturber or other mechanism to break the eTNO isotropy in the \des\ survey.  There is, however, a suggestive tendency for avoidance of $\Omega=0\degr$ in the \des\ sample, at 4--8\% significance, which motivates continuation of this exercise with \textit{LSST} or other future data.

\begin{figure}[h]
	\centering
	\includegraphics[width=\textwidth]{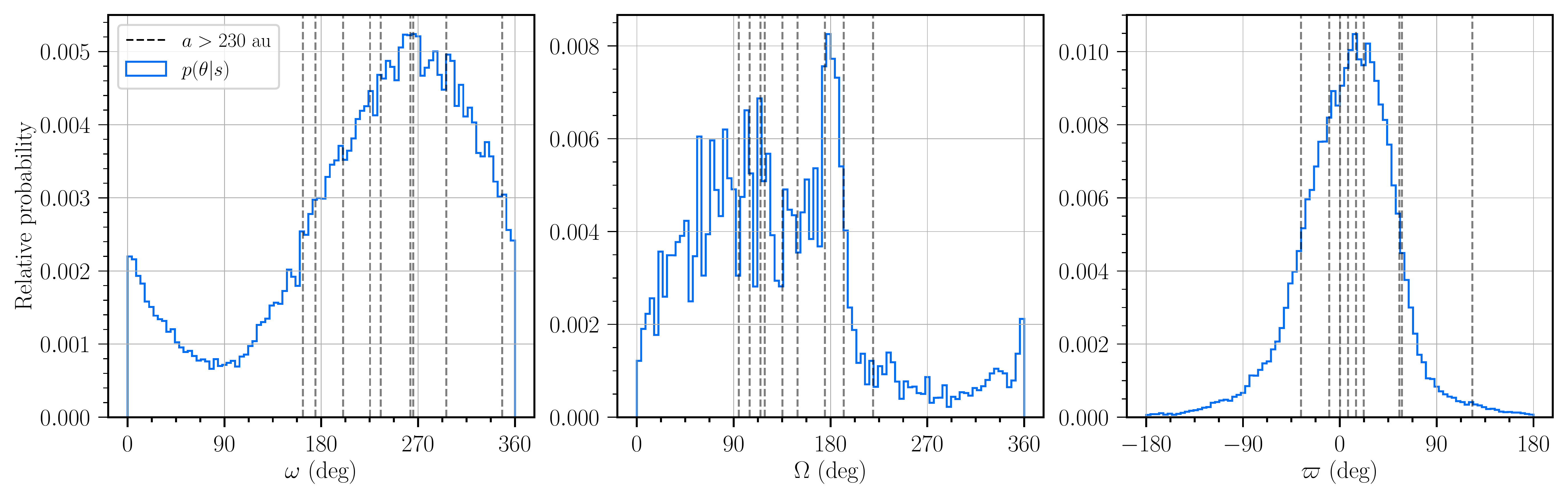}

	\caption{Histograms of the relative detection probabilities for the $a>230\au$ eTNOs (in blue) for $\omega$, $\Omega$ and $\varpi$ constructed as in Equation 1 of \cite{Bernardinelli2020}. The black vertical dashed lines represent the detected objects. \label{im:etno}}
\end{figure}

\subsection{Resonant dropouts}
We test the tendency of high-$q$ objects to be found preferentially ``sunward'' of distant resonances ($a$ lower than the nominal resonance center) rather than ``outward'' ($a$ higher than the nominal resonance center) \citep{Kaib2016,Nesvorny2016mig,PikeLawler2017}. {Such asymmetry is expected in models where ``resonance sweeping'' is carrying TNOs outward as Neptune migrates, but objects drop out of the resonance en route. This process is sensitive to the smoothness and speed of Neptune's migration \citep[see detailed discussion in][]{Kaib2016}. We conduct a similar test to the one presented in \cite{Lawler2019}, who found marginal inconsistency  ($p$-value $\approx 0.029$) of the OSSOS TNOs near the 5:2 and 4:1 resonances with $q>40\au$ arising from uniform distribution in period $P \equiv a^{3/2}.$}

Following the notation introduced in \cite{Bernardinelli2020}, we produce a model for the high-$q$ TNOs that is uniform in $\omega$, $\Omega$ and $\mathcal{M}$ and predicts a population $p(e,i,H|s)$ conditioned on a successful detection $s$ that reproduces the observed distribution in $\{e,i,H\}$:
\begin{equation}
	 p(a,e,i,\Omega,\omega,\mathcal{M},H) \propto \sum_j \frac{\delta(e - e_j) \delta (i - i_j) \delta (H - H_j)}{p(s|e_j,i_j,H_j)} u(\Omega) u(\omega) u(\mathcal{M}) p(a),
\end{equation}
where $\delta$ is the Dirac delta function, $u(\theta)$ is a uniform distribution in $\theta\in[0,2\pi)$, $p(a)$ is a chosen distribution in $a,$ and the sum is carried over the detected objects $j$. We simulate these detections following the procedure defined in Section \ref{sec:surveysim}, and obtain an ensemble of detected TNOs, where each object is weighted by the fraction of succesful detections $p(s|e,i,H)$.

We select the 16 non-resonant, non-scattering objects with $q > 38\au$ (so a larger sample can be obtained) within $\pm 2 \au$ of the 5:2 (2 objects, all sunward), 3:1 (7 objects, 5 sunward), 7:2 (3 objects, all sunward) and 4:1 (4 objects, all sunward) resonances, and test whether the sample is consistent with a distribution uniform in $P$ \citep[as in][]{Lawler2019}, as well as a distribution $p(a) \propto a^{-5/2}$ (as in the hot and stirred components of the CFEPS model). We define 
\begin{equation}
	\Delta P \equiv \frac{P - P_\mathrm{res}}{P_\mathrm{N}},
\end{equation}
{that is, the fractional difference between an object's period and the nominal resonance location in units of Neptune's period.
Objects with $\Delta P < 0$ ($\Delta P > 0$) are sunward (outward) of the resonance.
We apply a KS test between the observed distribution in $\Delta P$ and the distribution coming from the simulated ensemble with smooth $p(a)$, as shown in Figure \ref{im:resdrop}. The $p$-value of the KS statistic is determined by sampling $10^6$ different realizations of the smooth-$p(a)$ reference distribution. As an even simpler statistic, we calculate the expected fraction $f$ of the detected $\pm2$~au population expected to be sunward under the smooth-$p(a)$ reference model, and use binomial statistics to calculate the probability of matching or exceeding the 14 out of 16 real objects that are sunward.

\begin{figure}[h]
	\centering
	\includegraphics[width=0.49\textwidth]{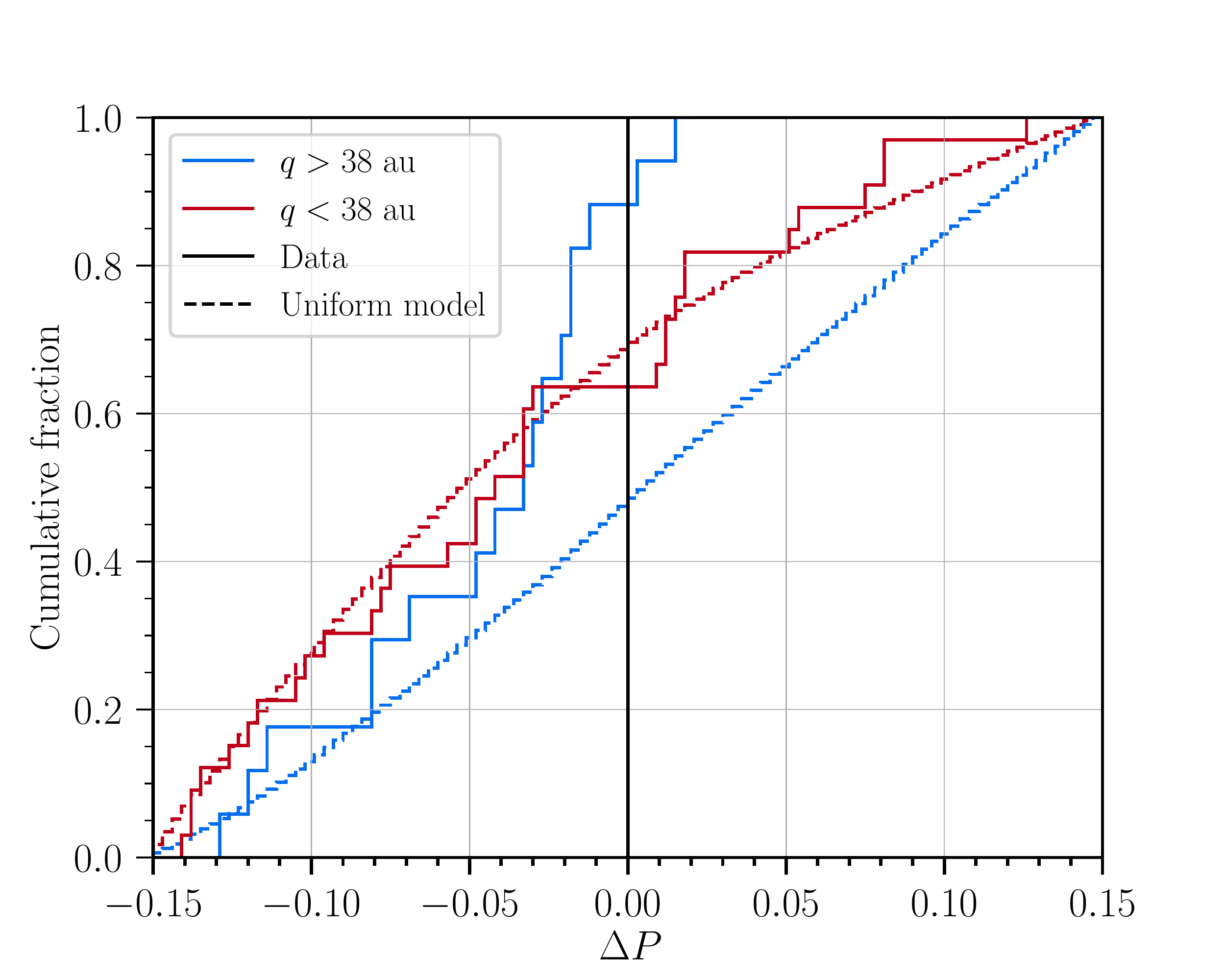}
	\includegraphics[width=0.49\textwidth]{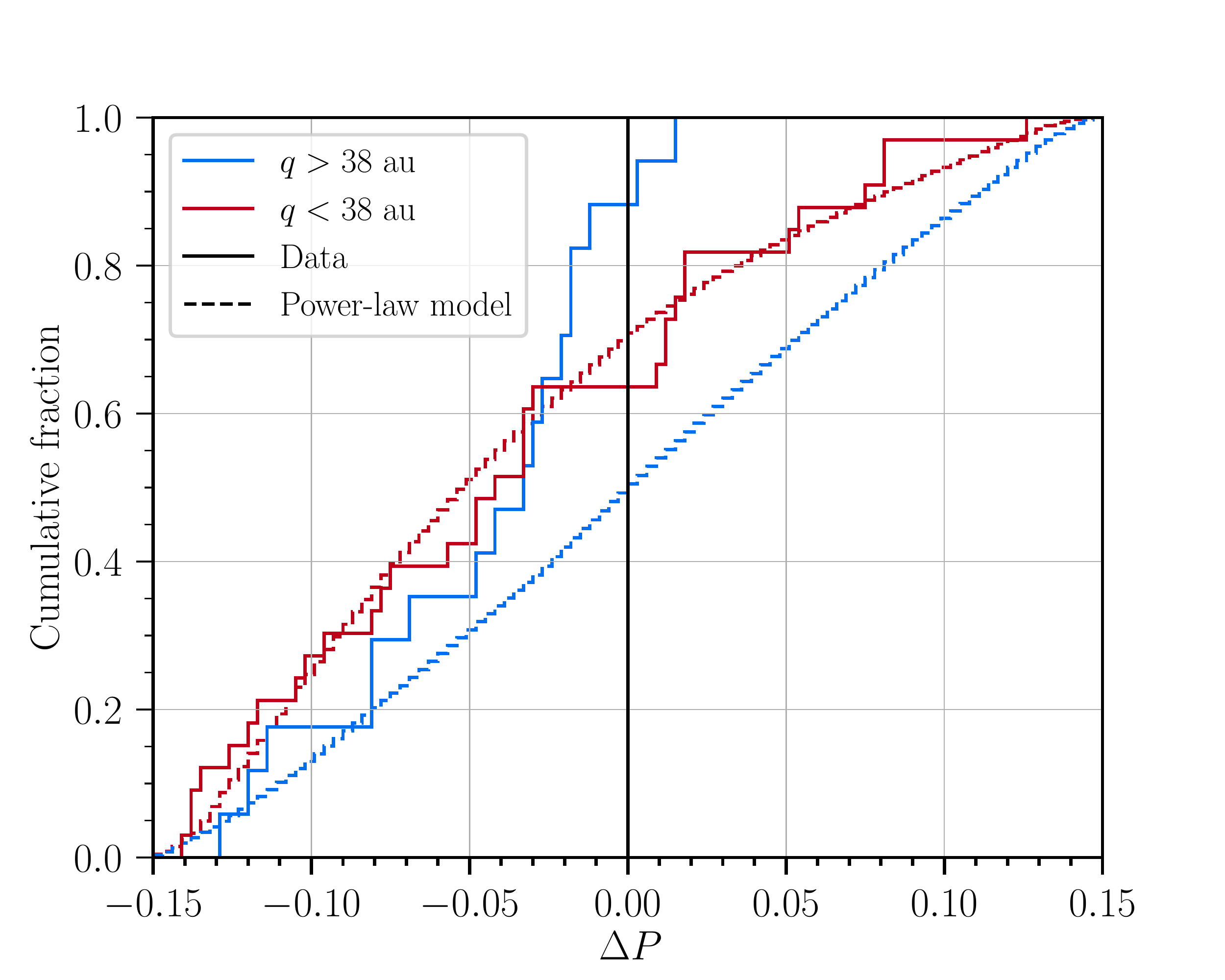}

	\caption{Selection functions in fractional period difference $\Delta P$ for the uniform (left) and $a^{-5/2}$ (right) models for the high (blue lines) and low (red lines) perihelion sample of objects near the $n$:1 and $n$:2 resonances. The solid lines show the cumulative fraction of detections in each population, and the dashed lines represent the detections expected from the model distribution. The solid black line divides the ``sunward'' ($\Delta P < 0$) from the ``outward'' ($\Delta P > 0$) detections. The distinction between these two groups is visually apparent, with the high-$q$ curves deviating significantly from the expected distribution under the assumed model.\label{im:resdrop}}
\end{figure}

Both the uniform-period and smooth power law $p(a)$ distributions can be rejected at very high significance for the first time, with $p=0.0003, 0.0007$ or $p=0.0008, 0.0010$ from the KS (indicating deviation from uniformity) and binomial (indicating a preference for sunward detections) tests, respectively, for this set of objects near the $n$:1 and $n$:2 resonances. 
We repeat these tests for the $q < 38\au$ non-resonant, non-scattering objects near the 5:2 (17 objects, 9 sunward), 3:1 (8 objects, all sunward), 7:2 (6 objects, 2 sunward) and 4:1 (2 objects, 0 sunward) resonances.  {Dynamical expectations are that perturbations from Neptune should weaken or erase any signature of resonant dropouts in this population \citep{Kaib2016}.  The 33 low-$q$ detections indeed show no evidence of preference for being sunward of resonance,  having $p$-values of 0.92 (uniform and smooth power law) and 0.58 (uniform), 0.69 (smooth power law) in the KS and binomial statistics, respectively.}

{These statistics confirm the clear visual impression from the upper panel of Figure~\ref{im:aei_y6} that in fact nearly all non-resonant TNOs beyond the 2:1 resonance with $q\gtrsim40$ are found just sunward of $n$:1 or $n$:2 resonances.}




 
\section{Summary}
\label{sec:summary}
We describe a catalog of 817 objects found in a complete search of the six-year, $5000\deg^2$ coverage Dark Energy Survey, yielding a sample 50\% complete at $m_r \approx 23.8.$ All detected objects have astrometry over multi-year arcs tied to Gaia DR2, and photometric measurements in the $grizY$ bands, with all measurements at the shot-noise limits. We demonstrate using both synthetic detections as well as previously discovered objects that the survey is highly complete given our thresholds of $\mathtt{NUNIQUE} \geq 7$, $\mathtt{ARCCUT} > 6$ months, and heliocentric distance $>29$~au. The \texttt{STS} technique discards virtually all accidental linkages, yielding a powerful methodology for discovery confirmation of faint sources. The Y6 sample is complete to $r_{50} = 23.77$, a gain of $\sim0.5$ mag over the Y4 sample, leading to the recovery of 506 objects that were not found in Y4, {all but 45 of which were new discoveries.}  We believe we have reached the limits of the \des\ data for detection of Solar System objects at $d>29\au$.
{A search for closer objects is currently beyond our computational means, as the burden grows as a high inverse power of $d$ and the search reported here required an estimated 15-20 million CPU hours. It could be made feasible by reducing the transient catalog density significantly, e.g. by imposing a minimum $S/N$.} This release includes only the $r$ band photometry---a future publication will describe our dedicated TNO photometric processing and report {detailed photometry for all bands and exposures.}

Searching for TNOs in the \des\ images presents challenges since it was not optimized for this purpose, such as its temporally sparse data, multi-band survey coverage, and extensive coverage at high ecliptic latitudes with lower TNO densities.  Despite this, the \des\ sample is comparable in number of objects to the largest predecessor TNO-targeted surveys. Its well characterized observational biases and publicly available survey simulator lead to a sample suited for statistical tests of Solar System population models. Indeed the fact the \des\ footprint is \emph{not} concentrated at the ecliptic will make the combination of the \des\ and OSSOS particularly powerful for constraining TNO population models.
The \des\ catalog represents a significant increase in the number of known dynamically detached objects, including new extreme and high-$q$ TNOs, as well as objects in distant resonances or high inclinations, which are relevant to more detailed hypothesis of the formation of the outer Solar System. 

The data set presented here also contains a number of new notable objects, in particular the Oort cloud comet C/2014 UN$_{271}$ (Bernardinelli-Bernstein), as well as the detached object \texttt{DES0502} with $i = 62.8\degr$ and the $q = 54.3\au$ object \texttt{DES0680}\footnote{These internal designations will be updated to MPC identifiers as soon as possible.}. 

Initial applications of  the statistical power of the \des\ Y6 TNO data are presented in three results for distinct populations of the trans-Neptunian region: 
\begin{itemize}
\item We show that our population of classical TNOs is formally incompatible with the CFEPS-L7 model for the $H_g < 8.5$ limit even if we change the relative sizes of the hot, stirred and kernel components. Such results indicate suggesting further substructure and/or a wider inclination distribution for the classical population.
\item The sample of extreme TNOs measured by \des\ is consistent with an underlying uniform distribution in $\omega$ and $\varpi$, agreeing with previous results from OSSOS \citep{Shankman2017}, \des, \citep{Bernardinelli2020} and their combination with the Sheppard-Trujillo surveys \citep{Napier2021}. There is a low significance disagreement between the observed $\Omega$ distribution and uniformity, and future surveys with large sky coverage, such as the upcoming \emph{Rubin Observatory's Legacy Survey of Space and Time} \citep{Ivezic2019}, will enable these two scenarios to be tested even further.
\item We measure with high significance the preference of high-$q$ objects to be found sunward of $n$:1 and $n$:2 resonances beyond 50~au, as expected from Solar System formation models with grainy or slow migration.  
\end{itemize}

The \des\ data, then, hold a unique set of objects comprising $\approx20\%$ of all currently-known TNOs, with well-characterized observational biases and covering $\approx 1/8$ of the sky.  These will be valuable for further detailed statistical tests of formation models for the trans-Neptunian region.

\appendix

\section{Appendix: Changes to the linking algorithm}
\label{sec:triplets}
We have updated the triplet finding algorithm presented in \citetalias{Bernardinelli2019}. Here, we will present only the details required to understand the changes, we refer the readers to \cite{Bernstein2000} and \citetalias{Bernardinelli2019} for the full details. 

Following the formalism of \cite{Bernstein2000}, we assume a coordinate system whose $z$ axis points to the center of a field that contains the object, with the origin being the location of the observatory at the midpoint the observing season ($t_0$). We use the six orbital parameters $\{\alpha = x/z,\beta=y/z,\gamma=1/z, \dot\alpha=\dot{x}/z, \dot\beta=\dot{y}/z,\dot\gamma = \dot{z}/{z}\}$. Under the approximation of inertial motion for the TNO (which is adequate for this purpose), the observed angular coordinates of the TNO at time $t$ are
\begin{equation}
	\boldsymbol\theta(t) = \frac{1}{1 + \dot\gamma (t - t_0) - \gamma z_\mathrm{obs}(t)}\left(\alpha + \dot\alpha (t - t_0) - \gamma x_\mathrm{obs}(t), \beta + \dot\beta (t - t_0) - \gamma y_\mathrm{obs}(t)  \right),
\end{equation}
Nominal orbital parameters for a given pair are determined by the 4 observed coordinates of the pair, plus a choice of 
nominal (inverse) distance $\gamma_0$ and an assignment $\dot\gamma_0=0$ to the line-of-sight velocity. For a given exposure being searched for a potential third detection, this defines a nominal search position $\boldsymbol\theta_0.$  We define two (non-orthogonal) vectors on the sky that track deviations of the predicted position as we vary the 2 degrees of freedom $\gamma$ and $\dot\gamma$ while keeping $\alpha$, $\beta$, $\dot\alpha$ and $\dot\beta$ constant:
\begin{align}
  \mathbf{u} & = \delta\gamma \frac{\partial\boldsymbol\theta}{\partial \gamma}, \\
  \mathbf{v} & = \dot\gamma_{\mathrm{bind}} \frac{\partial\boldsymbol\theta}{\partial \dot\gamma}.
\end{align}

These vectors span the maximum deviation of the pair's third detection from $\boldsymbol\theta_0$ if the TNO is assumed to lie in an (inverse) distance bin bounded by $\gamma_0\pm\delta\gamma$ and have line-of-sight velocity bounded by $|\dot\gamma| \le \dot\gamma_{\mathrm{bind}}.$  As the name suggests, $\dot\gamma_{\mathrm{bind}}$ is chosen to be the maximum allowed for a bound orbit given $\dot\alpha$, $\dot\beta$ and $\gamma_0$. 

Now the position of any potential third detection of the pair in this exposure can be written as
\begin{equation}
  \boldsymbol\theta = \boldsymbol\theta_0 + \mu \mathbf{u} + \nu \mathbf{v}, 
\end{equation}
 with $|\mu|<1, |\nu|<1.$ 
Conversely, each detection's position can be mapped to a pair $(\mu,\nu)$ by a linear transformation. To test whether a detection is inside the parallelogram defined by a bound orbit within the distance bin, it suffices to ask whether $|\mu|, |\nu| \le 1.$  This simple linear test leads to a speed gain of $100\times,$ on average, over the kD-tree triplet search of \citetalias{Bernardinelli2019}. Another benefit of the parallelogram search is a much simpler, parallelized implementation using \textsc{Numba} \citep{Numba}.
\newpage

\begin{acknowledgements}
\emph{Software}: The software developed in this work will be made public shortly after the publication. This work made use of the following public codes: \textsc{Numpy} \citep{Numpy}, \textsc{SciPy} \citep{SciPy}, \textsc{Astropy} \citep{Astropy2013,Astropy2018}, \textsc{Matplotlib} \citep{Matplotlib}, \textsc{IPython} \citep{iPython}, \textsc{easyaccess} \citep{Easyaccess}, \textsc{WCSFit} and \textsc{pixmappy} \citep{Bernstein2017astro}, \textsc{SExtractor} \citep{Bertin1996}, \textsc{CFITSIO} \citep{Cfitsio}, \textsc{Eigen} \citep{eigenweb}, \textsc{CSPICE} \citep{SPICE,ACTON20189}

University of Pennsylvania authors have been supported in this work by grants AST-1515804 and AST-2009210 from the National Science Foundation, and grant DE-SC0007901 from the Department of Energy.

Funding for the DES Projects has been provided by the U.S. Department of Energy, the U.S. National Science Foundation, the Ministry of Science and Education of Spain, 
the Science and Technology Facilities Council of the United Kingdom, the Higher Education Funding Council for England, the National Center for Supercomputing 
Applications at the University of Illinois at Urbana-Champaign, the Kavli Institute of Cosmological Physics at the University of Chicago, 
the Center for Cosmology and Astro-Particle Physics at the Ohio State University,
the Mitchell Institute for Fundamental Physics and Astronomy at Texas A\&M University, Financiadora de Estudos e Projetos, 
Funda{\c c}{\~a}o Carlos Chagas Filho de Amparo {\`a} Pesquisa do Estado do Rio de Janeiro, Conselho Nacional de Desenvolvimento Cient{\'i}fico e Tecnol{\'o}gico and 
the Minist{\'e}rio da Ci{\^e}ncia, Tecnologia e Inova{\c c}{\~a}o, the Deutsche Forschungsgemeinschaft and the Collaborating Institutions in the Dark Energy Survey. 

The Collaborating Institutions are Argonne National Laboratory, the University of California at Santa Cruz, the University of Cambridge, Centro de Investigaciones Energ{\'e}ticas, 
Medioambientales y Tecnol{\'o}gicas-Madrid, the University of Chicago, University College London, the DES-Brazil Consortium, the University of Edinburgh, 
the Eidgen{\"o}ssische Technische Hochschule (ETH) Z{\"u}rich, 
Fermi National Accelerator Laboratory, the University of Illinois at Urbana-Champaign, the Institut de Ci{\`e}ncies de l'Espai (IEEC/CSIC), 
the Institut de F{\'i}sica d'Altes Energies, Lawrence Berkeley National Laboratory, the Ludwig-Maximilians Universit{\"a}t M{\"u}nchen and the associated Excellence Cluster Universe, 
the University of Michigan, NSF's NOIRLab, the University of Nottingham, The Ohio State University, the University of Pennsylvania, the University of Portsmouth, 
SLAC National Accelerator Laboratory, Stanford University, the University of Sussex, Texas A\&M University, and the OzDES Membership Consortium.

Based in part on observations at Cerro Tololo Inter-American Observatory at NSF's NOIRLab (NOIRLab Prop. ID 2012B-0001; PI: J. Frieman), which is managed by the Association of Universities for Research in Astronomy (AURA) under a cooperative agreement with the National Science Foundation.

The DES data management system is supported by the National Science Foundation under Grant Numbers AST-1138766 and AST-1536171.
The DES participants from Spanish institutions are partially supported by MICINN under grants ESP2017-89838, PGC2018-094773, PGC2018-102021, SEV-2016-0588, SEV-2016-0597, and MDM-2015-0509, some of which include ERDF funds from the European Union. IFAE is partially funded by the CERCA program of the Generalitat de Catalunya.
Research leading to these results has received funding from the European Research
Council under the European Union's Seventh Framework Program (FP7/2007-2013) including ERC grant agreements 240672, 291329, and 306478.
We  acknowledge support from the Brazilian Instituto Nacional de Ci\^encia
e Tecnologia (INCT) do e-Universo (CNPq grant 465376/2014-2).

This manuscript has been authored by Fermi Research Alliance, LLC under Contract No. DE-AC02-07CH11359 with the U.S. Department of Energy, Office of Science, Office of High Energy Physics.



\end{acknowledgements}

\bibliography{references}
\bibliographystyle{aasjournal}

\allauthors
\end{document}